\documentclass[journal]{IEEEtran}
\usepackage{epsfig,graphics,graphicx,subfigure,amssymb,amstext,amsmath,algorithm,
algorithmic,wrapfig,multirow}
\usepackage{cite}
\usepackage{url}
\usepackage{times}
\usepackage{float}
\usepackage{placeins}
\usepackage{textcomp}
\usepackage[font=small,bf, skip=0pt]{caption}
\usepackage{hyperref}
\usepackage{flushend}
\usepackage{xcolor}
\usepackage{tabularx}
\usepackage{bm}
\usepackage{subfigure}

\def\beq{\begin{equation}}
\def\eeq{\end{equation}}
\def\beqa{\begin{eqnarray}}
\def\eeqa{\end{eqnarray}}
\def\beqan{\begin{eqnarray*}}
\def\eeqan{\end{eqnarray*}}

\def\C{{\mathbb{C}}}

\def\x{\times}

\def\rhat{\widehat{r}}

\def\tm1{t\! - \! 1}
\def\tp1{t\! + \! 1}

\def\dbf{\mathbf{d}}

\def\pbf{\mathbf{p}}

\def\rbf{\mathbf{r}}

\def\ubf{\mathbf{u}}

\def\vbf{\mathbf{v}}

\def\wbf{\mathbf{w}}

\def\xbf{\mathbf{x}}

\def\ybf{\mathbf{y}}
\def\zbf{\mathbf{z}}

\def\Abf{\mathbf{A}}

\def\Dbf{\mathbf{D}}

\def\Hbf{\mathbf{H}}
\def\Ibf{\mathbf{I}}

\def\Rbf{\mathbf{R}}

\def\Vbf{\mathbf{V}}

\def\Ybf{\mathbf{Y}}

\def\alphabf{{\boldsymbol \alpha}}

\def\betabf{{\boldsymbol \beta}}

\def\thetabf{{\boldsymbol \theta}}

\newif\ifconf
\conffalse

\newif\ifonecol
\onecolfalse

\IEEEoverridecommandlockouts
\renewcommand{\footnoterule}{%
  \kern -3pt
  \hrule width \columnwidth height 0.5pt
  \kern 3pt
}
\begin{document}

\title{Directional Cell Discovery in\\ Millimeter Wave Cellular Networks \vspace{-0.3cm}}

\ifconf
\else
    \author{
        C. Nicolas Barati~\IEEEmembership{Student Member,~IEEE},
        S. Amir Hosseini~\IEEEmembership{Student Member,~IEEE}, 
        Sundeep Rangan,~\IEEEmembership{Senior Member,~IEEE},
        Pei Liu~\IEEEmembership{Member,~IEEE}, \\
        Thanasis Korakis~\IEEEmembership{Senior Member,~IEEE},
        Shivendra S. Panwar,~\IEEEmembership{Fellow,~IEEE},
        Theodore S. Rappaport,~\IEEEmembership{Fellow,~IEEE}
        \thanks{This material is based upon work supported by the National Science
        Foundation under Grants No. 1116589 and 1237821 as well as generous support
        from Samsung, Nokia Siemens Networks, Intel, Qualcomm and InterDigital Communications.}
        \thanks{
            C. N. Barati (email: nicolas.barati@nyu.edu),
            S. Amir Hosseini (email: amirhs.hosseini@nyu.edu),
            S. Rangan (email: srangan@nyu.edu),
            T. Korakis (email: korakis@poly.edu),
            P. Liu (email: pliu@poly.edu),
            S. S. Panwar (email: panwar@catt.poly.edu),
            T. S. Rappaport (email: tsr@nyu.edu)
            are with the NYU WIRELESS Center,  Polytechnic School of Engineering,
            New York University, Brooklyn, NY.}
    }
\fi


\maketitle
\begin{abstract} The acute disparity between increasing bandwidth demand and available spectrum
has brought millimeter wave (mmWave) bands to the forefront of candidate solutions for the next-generation cellular networks.
Highly directional transmissions are essential for cellular communication in these frequencies to compensate for higher isotropic path loss.
This reliance on directional beamforming, however, complicates initial cell search since mobiles
and base stations must jointly search over a potentially large
angular directional space to locate a suitable path to initiate communication.
To address this problem, this paper proposes a directional cell discovery
procedure where base stations periodically transmit synchronization signals,
potentially in time-varying random directions, to scan the angular space.
Detectors for these signals are derived
based on a Generalized Likelihood Ratio Test (GLRT) under various signal and
receiver assumptions.
The detectors are then simulated under realistic design parameters
and channels based on actual experimental measurements at 28~GHz in New York City.
The study reveals two key findings:  (i) digital beamforming can
significantly outperform analog beamforming even when digital beamforming
uses very low quantization to compensate for the additional power requirements;
and (ii) omnidirectional transmissions of the synchronization signals
from the base station generally outperforms random directional scanning.
\end{abstract}


\section{Introduction}
\label{sec:intro}

Millimeter wave (mmWave) systems between 30 and 300~GHz have attracted considerable
recent attention for next-generation cellular
networks~\cite{KhanPi:11-CommMag,PietBRPC:12,rappaportmillimeter,RanRapE:14,andrews2014will, ghosh2014millimeter}.
The mmWave bands offer orders of magnitude more spectrum than current cellular allocations
-- up to 200 times by some estimates.
However, a key challenge in mmWave cellular is the signal range.
Due to Friis' Law~\cite{Rappaport:02}, the high frequencies of mmWave signals
result in large isotropic path loss (the free-space path loss grows with the frequency squared).
Fortunately, the small wavelengths of these signals also enable large number of antenna elements
to be placed in the same physical antenna area, thereby
providing high beamforming (BF) gains that can theoretically more than compensate
for the increase in isotropic path loss \cite{AkdenizCapacity:14, RapHe_mmWwc:14}.

However, for cellular systems, the reliance on highly directional transmissions
significantly complicates initial cell search.
While current cellular systems such as 3GPP LTE~\cite{3GPP36.300}
have considerable support for beamforming
and multi-antenna technologies, the underlying design assumption is that initial network
discovery can be conducted entirely with omnidirectional transmissions or transmissions
in fixed antenna patterns.
LTE base stations, for example, generally do not apply beamforming when
transmitting the synchronization and broadcast signals.
Adaptive beamforming and user-specific directional transmissions
are generally used only  \emph{after} the physical-layer access has been established.

However, in the mmWave range, it may be essential to exploit antenna gain even
during the cell search.  Otherwise,
the availability of high gain antennas would create a disparity between the range at which
a cell can be detected (before the correct beamforming directions are known
and the antenna gain is not available),
and the range at which reasonable data rates
can be achieved (after beamforming is used) -- a point made in ~\cite{LimmWBooster:13} and \cite[chapter 8]{RapHe_mmWwc:14}
and also illustrated in Fig.~\ref{fig:cellRange}.
This disparity would in turn create a large area where a mobile may potentially
be able to obtain a high data rate, but cannot realize this rate, since it cannot
even detect the base station (BS).

To understand this directional cell search problem,
this paper analyzes a standard cell search procedure where
the base station periodically transmits synchronization signals
and the mobiles scan for the presence of these signals to detect the base station,
and learn the timing and direction of arrivals.
This procedure is similar to the transmission of the Primary
Synchronization Signal (PSS) in LTE \cite{3GPP36.300}, except here we consider three
additional key
design questions specific to directional transmissions in the mmWave range:
\begin{itemize}
\item \emph{How should mobiles jointly search for base stations and
directions of arrival?}
The fundamental challenge for cell search in the mmWave range is the
directional uncertainty.  In addition to detecting the presence of
base stations and their timing as required in conventional cell search,
mmWave mobiles must also detect the spatial
angles of transmissions on which the synchronization signals are being received.

\item \emph{Should base stations transmit omnidirectionally or in
randomly varying directions?}
We compare two different strategies for the base station transmitter:
(i) the periodic
synchronization signals are beamformed in randomly varying directions to scan the angular space, and (ii) the signals are transmitted in a fixed omnidirectional pattern.  It is not obvious \emph{a priori}
which transmission strategy
is preferable:  randomly varying transmissions offers the possibility of occasional
very strong signals at the mobile receiver, while using an omnidirectional transmission allows a constant power at the receiver.
Note that, under uniform sampling of the angular directions,
the average power must be the same in both cases due to Friis' Law~\cite{Rappaport:02}.

\item \emph{What is the effect of analog vs.\ digital beamforming
at the mobile?}
Due to the high bandwidths and large number of antenna elements in the mmWave range,
it may not be possible from a power consumption perspective
for the mobile receiver to obtain high rate digital samples
from all antenna elements~\cite{KhanPi:11}.
Most proposed designs perform beamforming in analog (either in RF or IF)
prior to the A/D conversion
\cite{KohReb:07,KohReb:09,GuanHaHa:04,Heath:partialBF}.  A key limitation for these
architectures is that they permit the
mobile to ``look" in only one or a small number of directions at a time.
To investigate the effect of this constraint we consider two detection scenarios: (i) \emph{analog beamforming}
where the mobile beamforms in a random angular direction in each PSS time slot; and
(ii) \emph{digital beamforming}
where the mobile has access to digital samples from all the antenna elements. 
We will also consider  hybrid beamforming  \cite{Heath:partialBF,SunRap:cm14} -- see Section~\ref{hbf}.
\end{itemize}

The paper presents several contributions that may offer
insight to the above-listed questions.  First, to enable detection of the base stations
in the presence of directional uncertainty,
we derive generalized likelihood ratio test
(GLRT) detectors \cite{VanTrees:01a} that
treat the unknown spatial direction, delay and time-varying channel gains as unknown
parameters.
We show that for analog beamforming, the GLRT detection is equivalent to
matched filter detection with non-coherent combining across multiple PSS time slots.
For digital beamforming, the GLRT detector can be realized via
a vector correlation across all the antennas followed by a maximal eigenvector that finds the
optimal spatial direction.  Both detectors are computationally easy to implement.

Second, to understand the relative performance of omnidirectional
vs.\ random directions at the base station,
we simulate the GLRT detectors under realistic system parameters
and design requirements for both cases.
A detailed discussion of how the synchronization channel parameters
such as bandwidth, periodicity, and interval times should be selected
is also given.
The simulations indicate that, assuming GLRT detection,
omnidirectional BS transmissions
offer significant performance advantages in terms of detection time
in comparison to the base station randomly varying its transmission angles.
Of course, in either design,
once a base station is detected, further procedures would be needed
to more accurately determine and track the transmit and receive directions so that
subsequent communication can be directional -- we do not address this issue
in this paper.


Finally, the simulations also indicate that using analog beamforming
may result in a significant performance loss relative to digital beamforming.
For example, in the simulations we present, where the mobile has a 4$\times$4 uniform
2D array, the loss is as much as 18~dB.  Unfortunately, digital beamforming
requires separate A/D converters for each antenna, which may be
prohibitive from a power consumption standpoint in the bandwidths
used in the mmWave range.
However, the power consumption of the A/D can be dramatically reduced by
using very low bit rates (say 2 to 3 bits per antenna) as proposed by
\cite{Madhow:ADC,Madhow:largeArray} and related methods in \cite{hassan2010analog}. 
Using a standard white noise quantizer
 model~\cite{GershoG:92}, we show that, for the purpose of synchronization
 channel detection, the loss due to quantization noise
 with low bit rates is minimal.
We thus conclude that low-bit rate fully digital front-ends may be a superior design
choice in the mmWave range, at least for the purpose of cell search.

\ifonecol
\begin{figure}
\centering
\begin{minipage}{.45\textwidth}
  \centering
  \includegraphics[width=.78\linewidth]{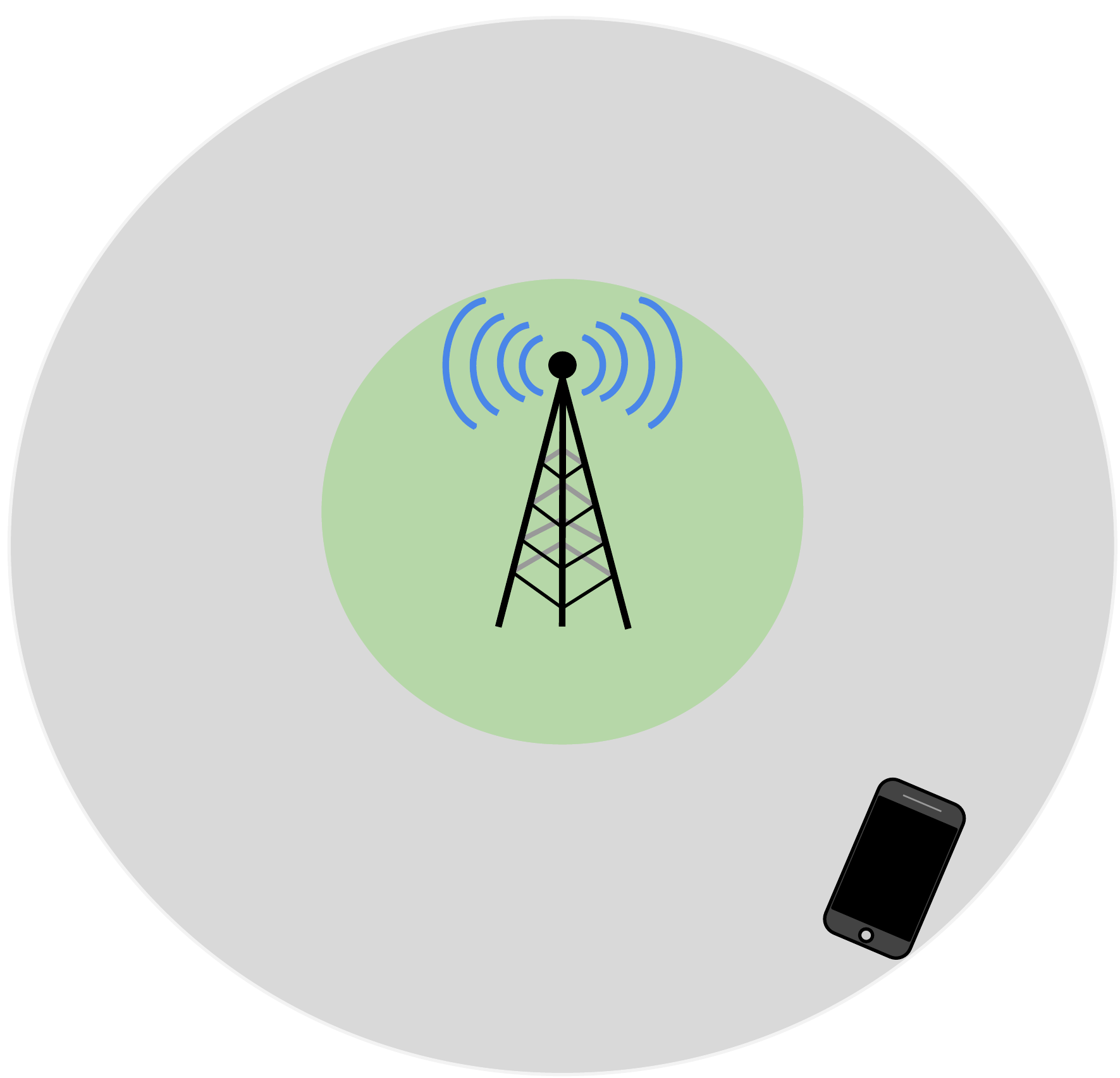}
  \captionof{figure}{The disparity in the
cell range in omnidirectional communication (green) vs.\ directional (gray).}
  \label{fig:cellRange}
\end{minipage}%
\begin{minipage}{.51\textwidth}
  \centering
  \includegraphics[width=1\linewidth]{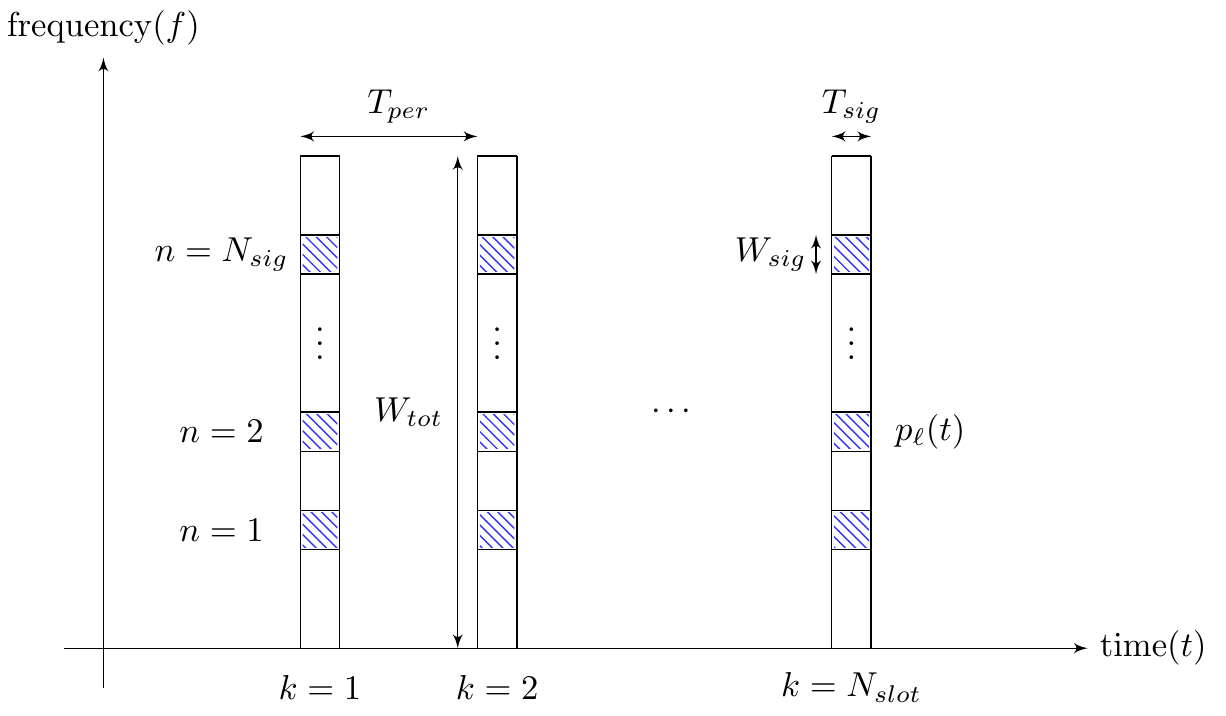}
  \captionof{figure}{Proposed PSS periodic transmission.}
  \label{fig:pssDesign}
\end{minipage}
\end{figure}

\else
\begin{figure}
\centering {\includegraphics[width=0.8\linewidth]{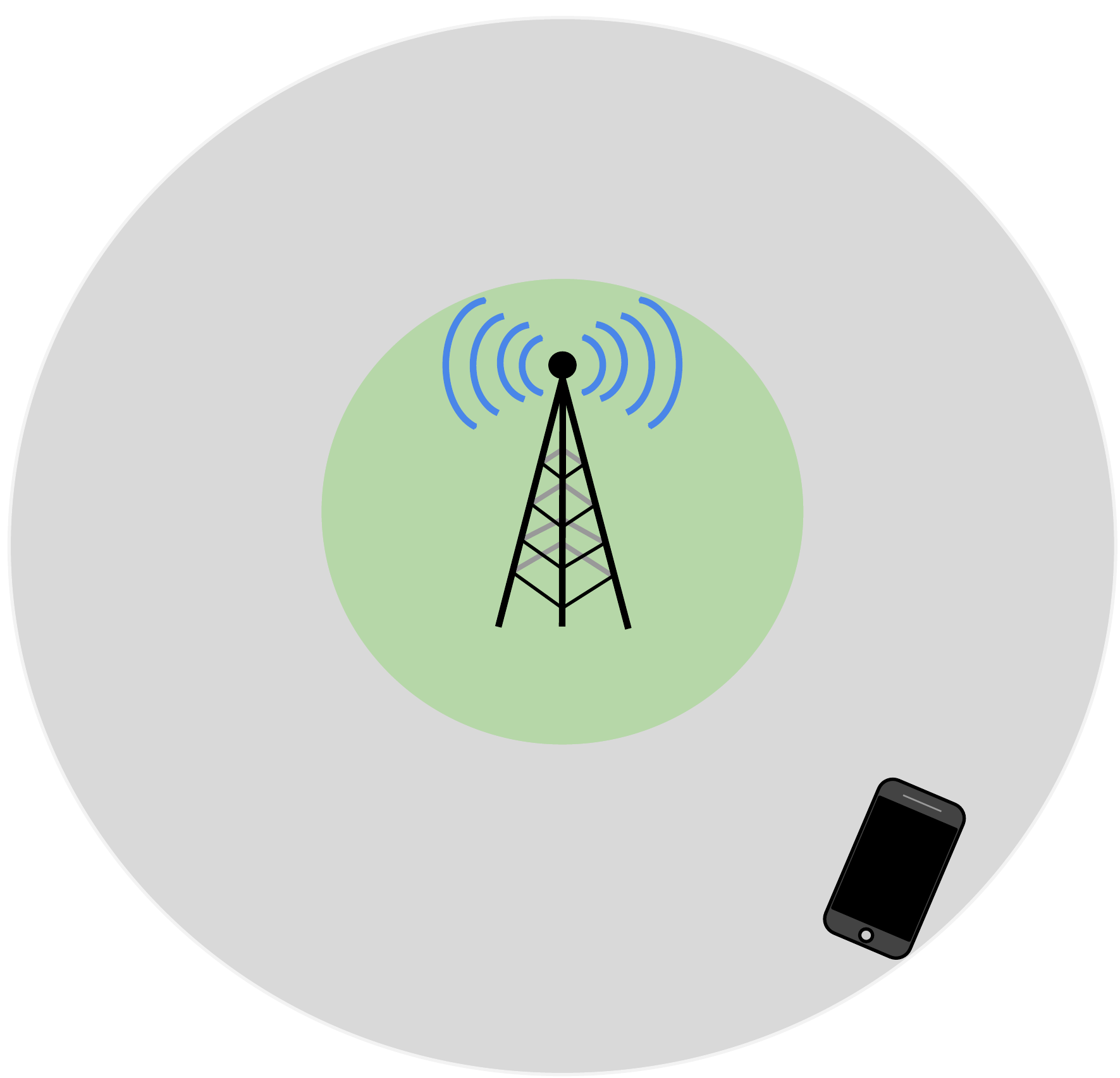}}
\caption{Due to highly directional transmissions,
millimeter wave (mmWave) systems will exhibit a large disparity in the
cell range in omnidirectional communication (green) vs.\ directional (gray).
If cell search does not exploit this antenna gain, mobiles in the outer gray
area may be capable of high data rates, but not able to locate the base station
to establish the communication.}
\label{fig:cellRange}
\end{figure}

\begin{figure}
\centering {\includegraphics[width=0.8\linewidth]{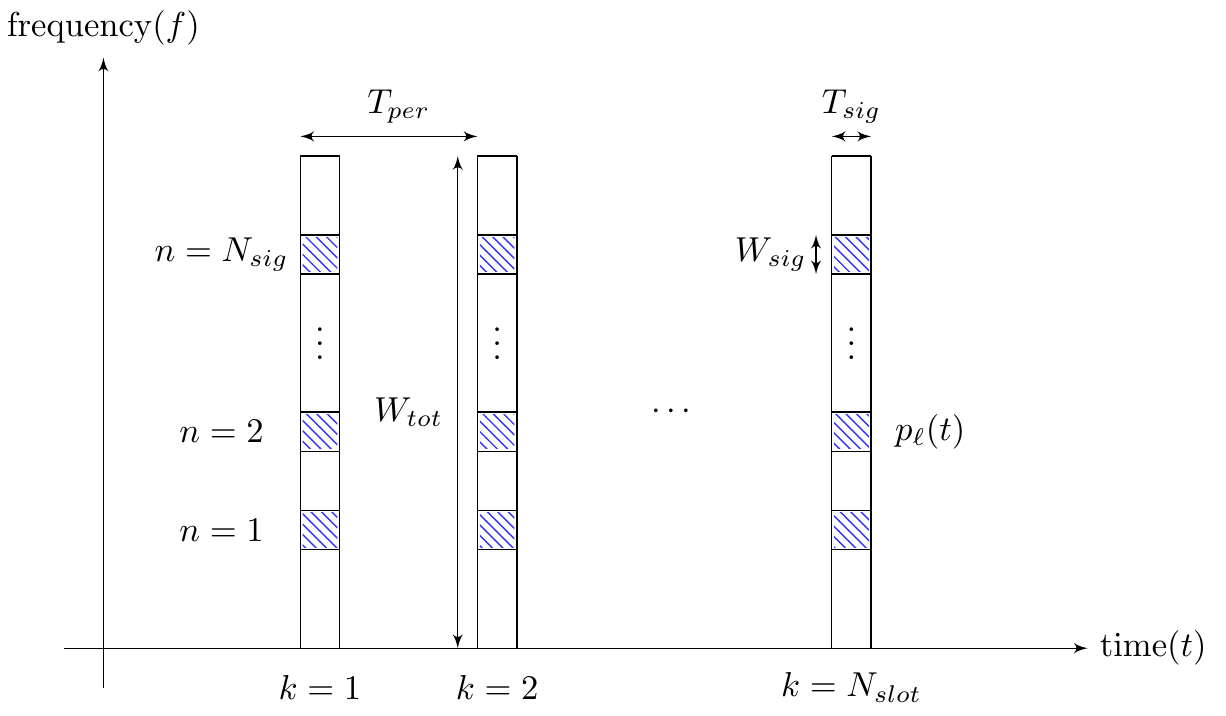}}
\caption{Proposed PSS periodic transmission.}
\label{fig:pssDesign}
\end{figure}
\fi
A conference version of this paper has appeared in \cite{BarHosCellSearch:14-spawc}.
This paper includes all the derivations, discussion of the parameter selection,
and more detailed, extensive simulations.

\section{Synchronization Signal Model} \label{sec:design}

\subsection{Synchronization Channel in 3GPP LTE}

We begin by briefly reviewing how synchronization channels are designed in LTE.
A complete description can be found in \cite{Dahlman:07}.
In the current LTE standard, each base station cell (called the evolved NodeB or eNB)
periodically broadcasts two signals:
the Primary Synchronization Signal (PSS) and the Secondary Synchronization Signal (SSS).
The mobiles (called the user equipment or UE)
search for the cells by scanning various frequency bands for
the presence of these signals.
The mobiles first search for the PSS which provides a coarse estimate of the frame timing,
frequency offset and receive power.  To simplify the detection, only one of three PSS signals
are transmitted.  Once the PSS is detected,
the UE can then search for the SSS.   Since the frame timing and frequency offset
of the eNB are already determined at this point, the SSS can belong to a larger set of 168 waveforms.
The index of the detected PSS and SSS waveforms together convey the eNB cell identity.
Since there are 3 PSS signals and 168 SSS signals, up to (168)(3)=504 cell identities can be communicated
with the two signal indices.
After the PSS and SSS are detected, the UE can proceed to decode the broadcast channels
in which further information of the base station can be read.

This cell search procedure is performed both by UEs in idle mode, either seeking
to establish an initial access to a base station or to find an eNB cell to ``camp on" for paging,
or UEs that are already connected to an eNB and need to look for other eNBs for potential handover targets.
After detecting an eNB through the cell search, UEs in idle mode
can initiate a random access procedure to the cell to establish a connection
to the eNB, should a transition to active mode be necessary.
UEs that are already
connected to a serving eNB can report the signal strengths of the detected cell IDs
to the network, which can then command the UE to perform a handover to that cell.

\subsection{Proposed mmWave Synchronization Channel Model}
In this work, we focus on the PSS design for mmWave systems, since this is the channel that
needs to be most significantly changed for mmWave.
We consider a PSS transmission scheme shown in Fig.~\ref{fig:pssDesign}.
Similar to the LTE PSS, we assume the signal is transmitted periodically once every
$T_{per}$ seconds in a brief interval of length $T_{sig}$.  We will call the short interval of
length $T_{sig}$ the \emph{PSS time slot}, and the period of length
$T_{per}$ between two PSS slots, the \emph{PSS period}.
The selection of $T_{sig}$, $T_{per}$  and other parameters will be discussed below.
As mentioned in the Introduction,
while LTE base stations generally transmit the PSS omnidirectionally or
in a fixed direction, here we consider two transmission scenarios:
(i) the base station transmits the synchronization signal omnidirectionally
(similar to current LTE), and
(ii) the base station randomly transmits the signal
in a different direction in each PSS time slot,
thereby randomly scanning the angular space. Deterministic search patterns, as in 802.11ad \cite{RapHe_mmWwc:14}, 
could also be performed although not considered here. 

Also, to exploit the higher bandwidths available in the mmWave range,
we assume that, in each PSS time slot,
the PSS waveform is transmitted over $N_{sig}$ PSS sub-signals
with each sub-signal being transmitted over a small bandwidth $W_{sig}$.
The use of multiple sub-signals
can provide frequency diversity, and narrowband signaling supports very low power receivers with high SNR capabilities.
In current LTE systems, the PSS signal is transmitted
over a relatively narrowband of approximately 930 kHz, which is slightly less than the
minimum possible system bandwidth of 1.08 MHz~\cite{Dahlman:07}.
Note that the PSS signal must fit in the minimum possible bandwidth.
A mobile searching for the base station does not know the actual bandwidth
during cell search, so all base stations transmit the PSS over the same bandwidth.
For mmWave systems, it is expected that the minimum bandwidth will be much larger
than current 4G systems, so that a reasonable number
of narrowband sub-signals can be accommodated, even in the minimum bandwidth.

\section{PSS Detector}

\subsection{Signal Model}

To understand the detection of the synchronization signal,
consider the transmission of a PSS signal from a BS with $N_{tx}$ antennas
to a mobile with $N_{rx}$ receive antennas.
Index all the PSS sub-signals across
all PSS time slots by $\ell$. Then,
we can write the complex baseband waveform for the PSS transmission from the BS as
\[
    \xbf(t) = \sum_{\ell=-\infty}^{\infty} \wbf^{tx}_\ell p_\ell(t),
\]
where $\xbf(t) \in \C^{N_{tx}}$ is the vector of complex signals to the $N_{tx}$ antennas,
$p_\ell(t)$ is the scalar PSS sub-signal waveform in the $\ell$-th time-frequency slot
and $\wbf^{tx}_\ell$ is the TX beamforming vector applied to the $\ell$-th sub-signal.
Note that in case of omnidirectional transmission, $\wbf^{tx}_\ell$ will be a constant
with all energy on a single antenna.
Since the sub-signals are transmitted periodically with $N_{sig}$ sub-signals per PSS time slot,
we will index the sub-signals so that
\beqa
     \lefteqn{ \mbox{support}(p_\ell) \subseteq I_k := [kT_{per},kT_{per}+T_{sig}] }, \nonumber \\
     && \ell \in J_k := \left\{ (k-1)N_{sig}+1, \ldots, kN_{sig}\right\}, \hspace{0.5cm}  \label{eq:pellSup}
\eeqa
where $I_k$ is the interval in time for the $k$-th PSS time slot, and $J_k$ is the set of indices $\ell$
such that the sub-signal $p_\ell(t)$ is transmitted in $I_k$.

For the purpose of the UE detector, our analysis is based on two key assumptions:  First,
we will assume that the channel is flat in the $T_{sig} \times W_{sig}$ time-frequency region
around each PSS sub-signal (Section~\ref{sec:simulation} will discuss the selection of the parameters
to ensure this).
Thus, the channel can be described by a sequence of
channel matrices $\Hbf_\ell \in \C^{N_{rx} \x N_{tx}}$ in case of directional transmission, and
$\Hbf_\ell \in \C^{N_{rx}}$ when the signal is transmitted omnidirectionally,
representing the complex channel gain
between the BS and mobile around the $\ell$-th PSS sub-signal.
Secondly, we will assume that the channel is rank one, which corresponds physically to a single
path with no angular dispersion.  In this case, the channel gain matrix can be written as
\[
    \Hbf_\ell = g_\ell \ubf\vbf^*,
\]
where $g_\ell$ is the small-scale fading of the channel in the $\ell$-th sub-signal
and $\ubf$ and $\vbf$ are the RX and TX spatial signatures, which are determined
by the large scale path directions and antenna patterns at the RX and TX.
With $\xbf^*$ or $\Abf^*$ we denote the conjugate transpose of any vector $\xbf$ or matrix $\Abf$.
Note that we use the rank one assumption only for deriving the detector -- our simulations in Section~\ref{sec:simulation}
will consider practical mmWave channels such as \cite{AkdenizCapacity:14,SunRap:cm14} with higher rank.
We assume that $\ubf$ and $\vbf$ do not change over the detection period
-- hence there is no dependence on $\ell$.
Of course, the real channel may not be exactly flat around the PSS sub-signals
and will also have some non-zero angular spread.
While our detector will assume a single path channel, our simulations
presented in Section~\ref{sec:simulation} consider a more accurate
multipath channel.

Under these channel assumptions,
the receiver will see a complex signal of the form,
\beq \label{eq:yrx}
    \ybf(t) = \sum_{\ell=-\infty}^{\infty}  \alpha_\ell \ubf p_\ell(t-\tau) + \dbf(t),
\eeq
where $\ybf(t)$ is the vector of RX signals across the $N_{rx}$ antennas,
$\tau$ is the delay of the PSS signal, $\alpha_\ell$ is the effective channel gain
\beq \label{eq:alphaell}
    \alpha_\ell = g_\ell \vbf^*\wbf^{tx}_\ell,
\eeq
and $\dbf(t)$ is AWGN.


\subsection{GLRT Detection with Digital RX Beamforming} \label{sec:GLRTdig}

The objective of the mobile receiver is
to determine, for each possible delay offset $\tau$,
whether a PSS signal is present at that delay or not.
Since the PSS is periodic with period $T_{per}$, the receiver needs only to test
delay hypotheses in the interval $\tau \in [0,T_{per}]$.

We first consider this PSS detection problem in the case when
the mobile can perform digital RX beamforming.  In this case, the mobile receiver
has access to digital samples from each of the individual
components of the vector $\ybf(t)$ in \eqref{eq:yrx}.
We assume the UE has access to full resolution values of the $\ybf(t)$.
We will address the issue of quantization noise -- critical for digital beamforming
-- in Section \ref{sec:simulation} and Appendix \ref{sec:quantLoss}.
We also assume that the detection is performed using $N_{slot}$ PSS time slots of data
\ifconf
\else
with indices $k=1,\ldots,N_{slot}$.  For a given delay candidate $\tau$,
let $\Ybf_{\tau}$ be the subset of the received signal $\ybf(t)$ for these time slots,
\beq \label{eq:Ytau}
    \Ybf_{\tau}  = \left\{ \ybf(t) ~\mid~ t \in I_k + \tau, ~ k=1,\ldots,N_{slot} \right\},
\eeq
where $I_k$ is defined in \eqref{eq:pellSup} and is the interval for the $k$-th PSS time slot.
\fi
Note that there are in total $L=N_{sig}N_{slot}$ sub-signals in this period -
each sub-signal indexed with $\ell$.

We can now pose the detection of the PSS signal as a binary hypothesis problem:
For each delay candidate $\tau$, a PSS signal is either present with that delay (the $H_1$
hypothesis) or is not present  (the $H_0$ hypothesis).  Following \eqref{eq:yrx},
we will assume a signal model for the two hypotheses of the form
\begin{subequations} \label{eq:ydigH01}
\beqa
    && H_1:  ~ \ybf(t) = \sum_{\ell=-\infty}^{\infty}  \alpha_\ell \ubf p_\ell(t-\tau) + \dbf(t), \\
    && H_0:  ~ \ybf(t) = \dbf(t),
\eeqa
\end{subequations}
where $\dbf(t)$ is complex white Gaussian noise with some power spectral density matrix $\nu \Ibf$.
Let $\thetabf$ be the vector of all the unknown parameters,
\[
    \thetabf = (\ubf,\nu,\alpha_1,\ldots,\alpha_L).
\]
\ifconf
\else
This vector
contains the unknown RX spatial direction, $\ubf$, the noise power density, $\nu$,
and the complex gains $\alpha_\ell$, $\ell=1,\ldots,L$.
\fi
Due to the presence of these unknown parameters, we use a Generalized
Likelihood Ratio Test (GLRT) \cite{VanTrees:01a}
to decide between the two hypotheses:
\beq \label{eq:GLRTDigCts}
    \Lambda(\tau) :=  \log
        \frac{\max_{\thetabf} p(\Ybf_\tau|H_1,\thetabf)}{
        \max_{\thetabf} p(\Ybf_\tau|H_0,\thetabf) }
      \underset{H_0}{\overset{H_1}{\gtrless}} \; t',
\eeq
\ifconf
\else
where $t'$ is a threshold and
\fi
$p(\Ybf_\tau|H_i,\thetabf)$ is the probability density of the received signal data $\Ybf_\tau$
under the hypothesis $H_0$ or $H_1$ and parameters $\thetabf$.
The GLR $\Lambda(\tau)$ is the ratio of the likelihoods
under the two hypotheses, maximized under the unknown parameters.
Note that although $\Ybf_\tau$ is a continuous signal, we assume that each PSS sub-signal lies
in a finite dimensional signal space.  Hence, the density is well-defined.
Also, one can in principle add other unknown parameters into the
channel model such as frequency errors due to Doppler shift or initial local
oscillator (LO) offsets.   However, it is computationally simpler to avoid adding
another parameter into the GLRT. Instead, as we discuss in Section \ref{sec:simulation},
we will account for frequency offset errors by discretizing
the frequency space and repeating the test over a number of frequency hypotheses.

\ifconf
\else
It is shown in Appendix~\ref{sec:glrtDig}
\fi
that the GLRT \eqref{eq:GLRTDigCts} can be evaluated
via a simple correlation.  Specifically, let $\vbf_\ell(\tau)$ be the
normalized matched-filter detector for the $\ell$-th subsignal,
\beq \label{eq:velldig}
    \vbf_\ell(\tau) = \frac{1}{\|p_\ell\|} \int p^*_\ell(t)\ybf(t-\tau)dt, \quad
    \|p_\ell\|^2 := \int |p(t)|^2dt.
\eeq
Also, let $E(\tau)$ be the total received energy in the $N_{slot}$ PSS time slots
under the delay $\tau$,
\beq \label{eq:EdigCts}
    E(\tau) = \sum_{k=1}^{N_{slot}} \int_{I_k} \|\ybf(t-\tau)\|^2 dt,
\eeq
where $I_k$ is the interval for the $k$-th time slot.
Then, it is shown that,
for any threshold level $t'$, there exists
a threshold level $t$ such that the GLRT in \eqref{eq:GLRTDigCts} is equivalent
to a test of the form,
\beq \label{eq:TDigRatCts}
    T(\tau) = \frac{\sigma_{max}^2(\Vbf(\tau))}{E(\tau)}
         \underset{H_0}{\overset{H_1}{\gtrless}} \; t
\eeq
where $t$ is a threshold level,
$\Vbf(\tau) \in \C^{N_{rx}\x L}$ is the matrix
\beq \label{eq:VdefCts}
    \Vbf(\tau) :=  \left[ \vbf_1(\tau), \cdots \vbf_L(\tau) \right],
\eeq
and $\sigma_{max}(\Vbf(\tau))$ is the largest singular value of the matrix.
The detector has a natural interpretation:  First,
we perform a matched filter correlator
for each sub-signal $p_\ell(t)$ on each of the vectors $\ybf(t)$ yielding a
vector correlation output $\vbf_\ell(\tau)$.
We then compute the maximum singular vector
for the matrix $\Vbf(\tau)$ which finds the energy in the most likely spatial direction
across all the $L$ sub-signals.
Note that the GLRT is invariant to scaling of the
received signal $\ybf(t)$ since the channel gain is treated as unknown.

It should also be noted that finding the maximum singular vector does add some complexity
relative to PSS searching in LTE which only needs matched filtering.
However, finding a maximum singular vector can be performed relatively easily,
for example, via a power method~\cite{GolubV:89}.  Thus, we anticipate that the
computation should be feasible, particularly as other parts of the baseband processing
would need to scale for the higher data rates.

\subsection{GLRT for Analog Beamforming} \label{sec:GLRTanalog}

We next consider the hypothesis testing problem in the case when the mobile
receiver can only perform beamforming in analog (either at RF or IF).  In this case,
for each PSS time slot $k$, the RX must select a receive beamforming vector
$\wbf^{rx}_k \in \C^{N_{rx}}$ and then only observes the samples in this direction.
If we let $z(t)$ be the scalar output from applying the beamforming vector $\wbf^{rx}_k$
to the received vector $\ybf(t)$, the signal model for the two hypotheses in
\eqref{eq:ydigH01} is transformed to
\begin{subequations} \label{eq:zH01}
\beqa
    && H_1:  ~ z(t) = \sum_{\ell=-\infty}^{\infty}  \beta_\ell  p_\ell(t-\tau) + d(t), \\
    && H_0:  ~ z(t) = d(t),
\eeqa
\end{subequations}
where $\beta_\ell$ is the effective gain after TX and RX beamforming and
$d(t)$ is complex white Gaussian noise with some PSD $\nu$.  From \eqref{eq:alphaell},
the beamforming gain will be given by
\beq \label{eq:hell}
    \beta_\ell :=  g_\ell {\wbf^{rx}_k}^*\ubf\vbf^*\wbf^{tx}_k,
\eeq
\ifconf
\else
for all sub-signals $\ell$ in the time slot $k$ (i.e.\ $\ell \in J_k$).
\fi
Here, we have assumed that both the TX and RX must apply the same beamforming gains ($1$ in omni case)
for all sub-signals in the same PSS time slot.  This requirement is necessary since, under analog
beamforming, both the TX and RX can use only one beam direction at a time.
The unknown parameters in the analog case can be described by the vector
\ifconf
\else
\[
    \thetabf = (\nu, \beta_1,\ldots,\beta_L),
\]
which contains the unknown variance $\nu$ and the channel gains $\beta_1,\ldots,\beta_L$.
\fi

Analogous to \eqref{eq:GLRTDigCts}, we use a GLRT of the form
\beq \label{eq:GLRTAnalogCts}
    \Lambda(\tau) :=  \log
        \frac{\max_{\thetabf} p(Z_\tau|H_1,\thetabf)}{
        \max_{\thetabf} p(Z_\tau|H_0,\thetabf) }
     \underset{H_0}{\overset{H_1}{\gtrless}} \; t',
\eeq
where $t'$ is a threshold level and $Z_{\tau}$
\ifconf
\else
is defined similarly to \eqref{eq:Ytau} for the beamformed signal $z(t)$.
\fi

Similar to the digital case, this GLRT can be evaluated via a correlation.
Let $v_\ell(\tau)$ be the normalized correlation of $z(t)$ with the sub-signal $p_\ell(t)$,
\beq \label{eq:vellanalog}
    v_\ell(\tau) = \frac{1}{\|p_\ell\|} \int p^*_\ell(t)z(t-\tau)dt, \quad
    \|p_\ell\|^2 := \int |p(t)|^2dt.
\eeq
It is shown
\ifconf
\else
in Appendix~\ref{sec:glrtAnalog}
\fi
that GLRT \eqref{eq:GLRTAnalogCts}
is equivalent to a test of the form
\beq \label{eq:TRatAnalogCts}
    T(\tau) = \frac{\|\vbf(\tau)\|^2}{E(\tau)}
    \underset{H_0}{\overset{H_1}{\gtrless}} \; t,
\eeq
where $t$ is a threshold level and $\vbf(\tau) \in \C^{L}$ is the vector
\beq \label{eq:vdefAnalogCts}
    \vbf(\tau) :=  \left[ v_1(\tau), \cdots, v_L(\tau) \right],
\eeq
and $E(\tau)$ is the energy in the $N_{slot}$ time slots,
\beq \label{eq:EanalogCts}
    E(\tau) = \sum_{k=1}^{N_{slot}} \int_{I_k} \|z(t-\tau)\|^2 dt.
\eeq
Thus, in the analog beamforming case, the GLRT is simply performed with non-coherently adding the
energy from the matched filter outputs for the $L$ sub-signals.
An identical detector can also be used for hybrid beamforming where the outputs from multiple streams
are treated as separate measurements -- see Section~\ref{hbf}.

\section{Simulation} \label{sec:simulation}


We assess the performance of the directional correlation detector
for three cases: analog, digital, and quasi-analog or hybrid  BF.
We provide some explanation regarding the latter at the end of this section.
Although we have derived our detectors based on a flat single path channel,
we simulate the detectors' performance both for a single path channel,
as well as multipath pathloss channel model \cite{AkdenizCapacity:14} 
derived from actual measurements in New York City~\cite{Rappaport:12-28G,Rappaport:28NYCPenetrationLoss,Samimi:AoAD,rappaportmillimeter,McCartRapICC15}.
The parameters for the channels (Table \ref{tab:simPar})
are based on realistic system design considerations.
A detailed discussion of the selection of these parameters is given in
Appendix~\ref{sec:simParam}.
Briefly, the PSS parameters $T_{sig}$ and $W_{sig}$ and $T_{per}$
were selected to ensure the channel is roughly flat within each $T_{sig} \x W_{sig}$ PSS sub-signal
time-frequency region based on typical time and frequency coherence bandwidths observed in
\cite{Rappaport:12-28G,McCartRapICC15}.
The parameter $T_{sig}$ was also selected sufficiently short so that
the Doppler shift across the sub-signal would not be significant
with moderate mobile velocities (30~km/h).
The value $T_{per}$ was then selected to
keep a low (2\%) total PSS overhead ($=T_{sig}/T_{per}$).
This is slightly above the 1.5\% overhead of PSS in LTE.
We assume that during each PSS slot \emph{only} the synchronization signal is transmitted and nothing else.
The value $N_{sig}=4$ was found by trial-and-error to give the best performance in terms of frequency diversity versus energy loss from
non-coherent combining.  Under these parameters, we selected $N_{slot} = 50$ slots to perform the search,
which corresponds to a search time of $N_{slot}T_{per}=$ $250$ ms --
a reasonable time frame for initial access.

For the antenna arrays, we considered 2D uniform linear arrays with $4 \times 4$
elements at the mobile, and $8 \times 8$ at the base station.  When needed,
random directions were selected by applying the beamforming for a randomly selected
horizontal  and elevation angle.  Note that this selection only requires random phases
applied to the antenna elements -- not magnitude.

To compute the threshold level $t$ in \eqref{eq:TDigRatCts} and
\eqref{eq:TRatAnalogCts},
we first computed a false alarm probability target with the formula
\[
    P_{FA} = \frac{R_{FA}}{N_{PSS}N_{dly}N_{FO}},
\]
where $R_{FA}$ is the maximum false alarm rate per search period over
all signal, delay and frequency offset hypotheses and $N_{PSS}$, $N_{dly}$
and $N_{FO}$ are, respectively, the number of signal, delay and frequency offset
hypotheses.  Again, the details of the selection are given
in Appendix~\ref{sec:simParam}. The delay hypotheses $N_{dly}$
were calculated assuming sampling at twice the PSS bandwidth
so that $N_{dly} = 2W_{sig}T_{per}$.
The number of frequency offsets $N_{FO}$ hypotheses was computed
so that the frequency search could cover both an
initial local oscillator (LO) error of $\pm 1$~part per million (ppm) as well as Doppler shift
up to 30~km/h at 28~GHz.
We selected
$N_{PSS}=3$ for the number of signal hypotheses as used in current 3GPP LTE.
We then used a large number of Monte Carlo trials to find the
threshold to meet the false alarm rate.
Since $P_{FA}$ was very small, we extrapolated the tail
distribution of the statistic to estimate the correct threshold analytically,
assuming $\log \Pr(T > t)$ is quadratic in $t$ for large $t$.

\subsection{Detection Performance with Single Path Omnidirectional Transmissions}

\begin{figure}
\centering {\includegraphics[width=0.8\linewidth]{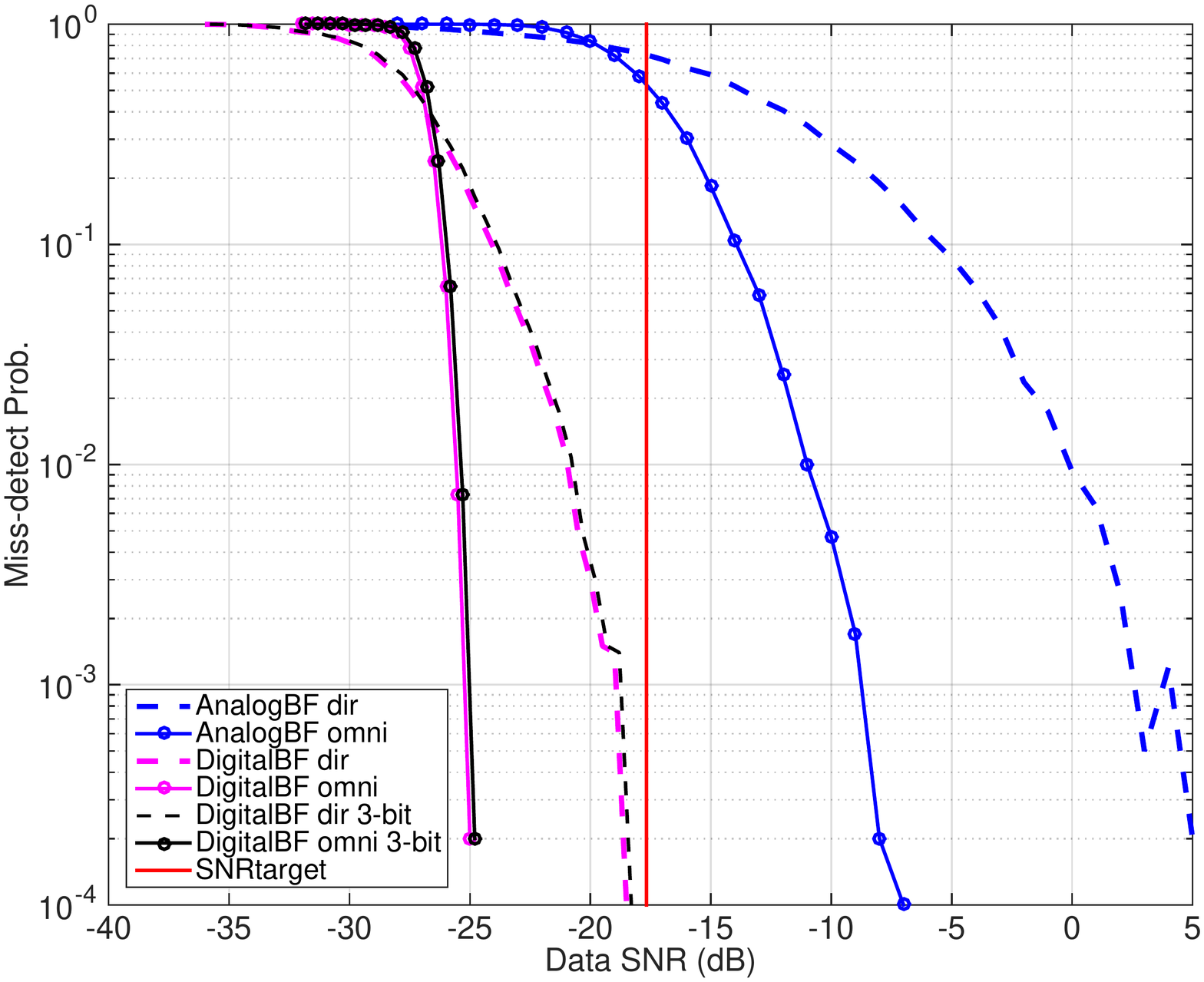}}
\caption{Misdetection Probability vs.\ data SNR: base station transmits PSS omnidirectionally (solid), and  in random directions (dashed). In both cases, the performance
of analog (blue) and digital (magenta) beamforming are compared at the receiver end.
Also, a quantized version of digital beamforming (black) shows the small loss due to low bit rate (3 bits) resolution.
Target SNR for $R_{tgt}=$ 10 Mbps (red).}
\label{fig:DvO}
\end{figure}
\begin{figure}
\centering {\includegraphics[width=0.8\linewidth]{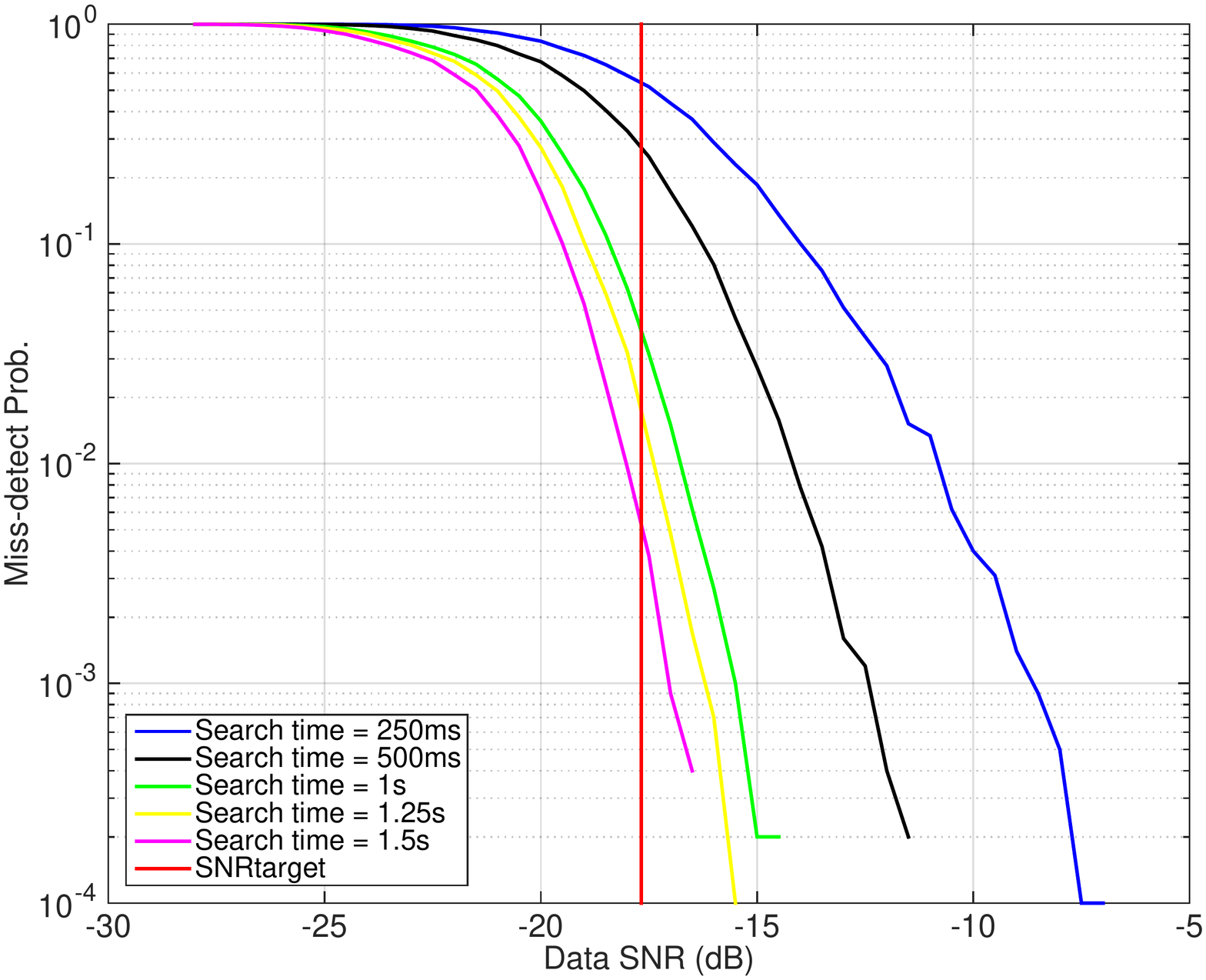}}
\caption{Misdetection Probability vs.\ data SNR: Omnidirectional transmission.  The effect of
increasing the search time while keeping the overhead fixed at $2\%$.
Target SNR for $R_{tgt}=$ 10 Mbps (red)}
\label{fig:OH}
\end{figure}

First, we assessed the performance of the GLRT
detectors in the case where the synchronization
signal is transmitted omnidirectionally every $T_{per}$ seconds.
The result is summarized in Fig.~\ref{fig:DvO}.
The figure shows a large  gap in SNR between digital and analog beamforming
--  more than 20 dB for $P_{MD}=0.01$.
This gap is largely due to the fact that digital beamforming with the proposed eigenvector
correlator can, in essence, determine the correct spatial direction over all the sub-signals,
while analog beamforming can only ``look" in one direction at a time.

The definition of SNR in Figures~\ref{fig:DvO} through~\ref{fig:hbf}, requires some explanation.
The SNR that determines the performance is what we will call the
\emph{PSS SNR} given by
\beq \label{eq:snrPss}
    \mbox{SNR}_{\rm PSS} = P T_{sig}/(N_0N_{sig}),
\eeq
which is the SNR on a single PSS sub-signal for an omnidirectional received power $P$
and noise density $N_0$.
However, the PSS SNR has no meaning outside the PSS signal.
So, in Figures~\ref{fig:DvO} through~\ref{fig:hbf}, we instead plot the misdetection rate against what
we call the \emph{data SNR} given by
\beq \label{eq:snrData}
        \mbox{SNR}_{\rm data} = P G_{tx,max}  G_{rx,max} / (N_0 W_{tot}),
\eeq
where $G_{tx,max}$ and $G_{rx,max}$
are the maximum possible beamforming gains and
$W_{tot}$ is the total available bandwidth.  The data SNR \eqref{eq:snrData}
is the theoretical SNR for a signal transmitted across the entire bandwidth $W_{tot}$
assuming optimal beamforming.
The data SNR is related to the PSS SNR by
\beq \label{eq:pssDataSnr}
    \mbox{SNR}_{\rm PSS} = \mbox{SNR}_{\rm data}\frac{T_{sig}W_{tot}}
    {N_{sig}G_{tx,max}G_{rx,max}}.
\eeq
So, we simulate the PSS detection at the PSS SNR, but plot the result
against the data SNR.

When plotted against the data SNR,
it is easy to interpret the results of Fig.~\ref{fig:DvO}.
For example, consider a typical edge rate minimum SNR requirement.
Suppose that the UE should be able to
connect whenever the SNR is sufficiently large to support some target rate, $R_{tgt}$.
If the system was operating at the Shannon capacity with optimal beamforming,
it would achieve this rate whenever the data SNR satisfies,
\[
    R_{tgt} = \beta W_{tot}\log_2\left(1 + \mbox{SNR}_{\rm data}\right),
\]
where $\beta$ is a bandwidth overhead fraction.
Fig.~\ref{fig:DvO} shows the target data SNR
for $R_{tgt}$=10 Mbps, $W_{tot} = 1$ GHz and $\beta = (0.5)(0.8)$ to account for half-duplexing
TDD constraints and 20\% control overhead \cite{AkdenizCapacity:14}.
From Fig.~\ref{fig:DvO}, we see that with digital beamforming, the system can reliably detect the
signal at the SNR for the 10 Mbps rate target.
But, with analog beamforming, the system needs a significantly
larger SNR.

This edge rate requirement analysis also illustrates another issue.
An edge rate target determines a minimum data SNR, which in turn
determines the minimum PSS SNR.  From \eqref{eq:pssDataSnr}, we see that for a given
data SNR requirement, the PSS SNR decreases with the beamforming gain.
In mmWave systems, this beamforming gain can be large.  For example,
in a single path channel, the beamforming gain is equal to the number of antennas.
So, in our simulations,  $G_{tx,max} = 64$ and $G_{rx,max}=16$ for a
combined total gain of 30~dBi.
Gains of more than 20~dBi  per side are not common in practical systems~\cite{ghosh2014millimeter}.
It is precisely this gain that makes the synchronization a challenge for mmWave systems:
high antenna gains imply that the link can meet minimum rate requirements at very low SNRs once the link is
established and directions are discovered.  But, in order
to use the links in the first place,
the mobiles need to be able to detect the base stations at these low SNRs before the directions are known.

Note that the edge rate target would likely be determined by the point at which it is better
to shift the mobiles from the 4G system.  If a 100~Mbps edge rate were used,
as suggested in \cite{andrews2014will,ghosh2014millimeter}, the SNR target for which
the mobile must detect the mmWave base station would be higher, making the detection
requirement easier.  However, using this target would imply less mobiles can be supported by 5G.
Similarly, if the mmWave base station has a higher antenna gain, the PSS detection rate would
be lower for the
same data SNR.  For example, if the base station had 256 antennas instead of 64 as assumed in this
simulation, the data SNR would need to increase by 6~dB to maintain the same misdetection rate.
Compensating for this SNR loss would necessitate longer detection times or more overhead.

Finally, given the results in Fig.~\ref{fig:DvO},
we wish to see if there is some way to improve
analog beamforming's performance.
Fig.~\ref{fig:OH} shows that increasing the total search time by non-coherent
combining over
larger numbers of slots can lead to an improvement.  However, the performance
improvements show diminishing returns with increasing search time, and such long search times
may induce intolerable delay overhead.
Alternatively, we could keep the search time fixed and increase the overhead
by transmitting the PSS signals more  frequently.
Either way, if we are to insist on an analog detector,
sacrifices in searching time and possibly overhead may be necessary.

\begin{table}
\begin{center}
\ifonecol
\begin{tabular}{|p{6 cm}|p{5 cm}|}
\else
\begin{tabular}{|p{3.25 cm}|p{3 cm}|}
\fi
  \hline
  {\bf Parameter} & {\bf Value}
  \tabularnewline \hline

Total system bandwidth, $W_{tot}$ & 1 GHz
\tabularnewline \hline

Number of sub-signals per PSS time slot, $N_{sig}$    & $4$
\tabularnewline \hline

Subsignal Duration, $T_{sig}$ & $100$  $\mu$s
\tabularnewline \hline

Subsignal Bandwidth, $T_{sig}$ & 1 MHz
\tabularnewline \hline

Period between PSS transmissions, $T_{per}$  & 5 ms
\tabularnewline \hline

PSS overhead & 2\%
\tabularnewline \hline

Search period, $N_{slot}$ & 50 slots = 250~ms
\tabularnewline \hline

Total false alarm rate per search period,
$R_{FA}$ & $ 0.01$
\tabularnewline \hline

Number of PSS waveform hypotheses, $N_{PSS}$  & $3$ \tabularnewline \hline
Number of frequency offset hypotheses, $N_{FO}$ & $ 23 $ \tabularnewline \hline

BS antenna & $8\times8$ uniform linear array   \tabularnewline \hline
UE antenna & $4\times4$ uniform linear array   \tabularnewline \hline

Carrier Frequency & 28 GHz \tabularnewline \hline

PSS SNR, $PT_{sig}/N_0$ & varied \tabularnewline \hline
\end{tabular}
\caption{Default simulation parameters unless otherwise stated.}
\label{tab:simPar}
\end{center}
\end{table}

\subsection{Detection Performance with a Single Path Channel and
Randomly Varying Transmission Angles}

We next consider the case where the synchronization signal is transmitted in a random
direction at every transmission instant.
A deterministic search pattern, such as in 802.11ad \cite{RapHe_mmWwc:14} could also be used, but is not considered in this work.
The simulation parameters remain the same as
before and the results are also summarized in Fig.~\ref{fig:DvO}.

By switching to directional PSS transmission, the detector's performance seems to degrade significantly: about $11$~dB in the analog case and $5$~dB with digital beamforming.
In this case, as mentioned in Section~\ref{sec:GLRTanalog} the analog detector sums up the match-filter's output
non-coherently for every given delay offset $\tau$.
The reason for the degradation is that, when the base station transmits in random
directions, the beams often ``miss" the UE and the UE sees very little signal energy.
In the omni case, on the other hand, although the signal loses about $18$~dB of potential TX beamforming gain,
there is at least a constant (weak) power reception.
The results are similar in other parameter settings, and we conclude that, for synchronization signals,
randomly scanning angles at the transmitter is worse, in general,
than using a constant omnidirectional transmission.
It is worth noting that realistic omnidirectional  mmWave antennas at the BS transmitter 
will likely use downtilt with significant gain on or below the horizon, just as is done in today's cellular systems.
In this case, some of the relative benefit of omnidirectional transmission over randomly varying directions
will be reduced.

\subsection{Detection Performance with Multipath Channels from New York City Data}
\begin{figure}
\centering {\includegraphics[width=0.8\linewidth]{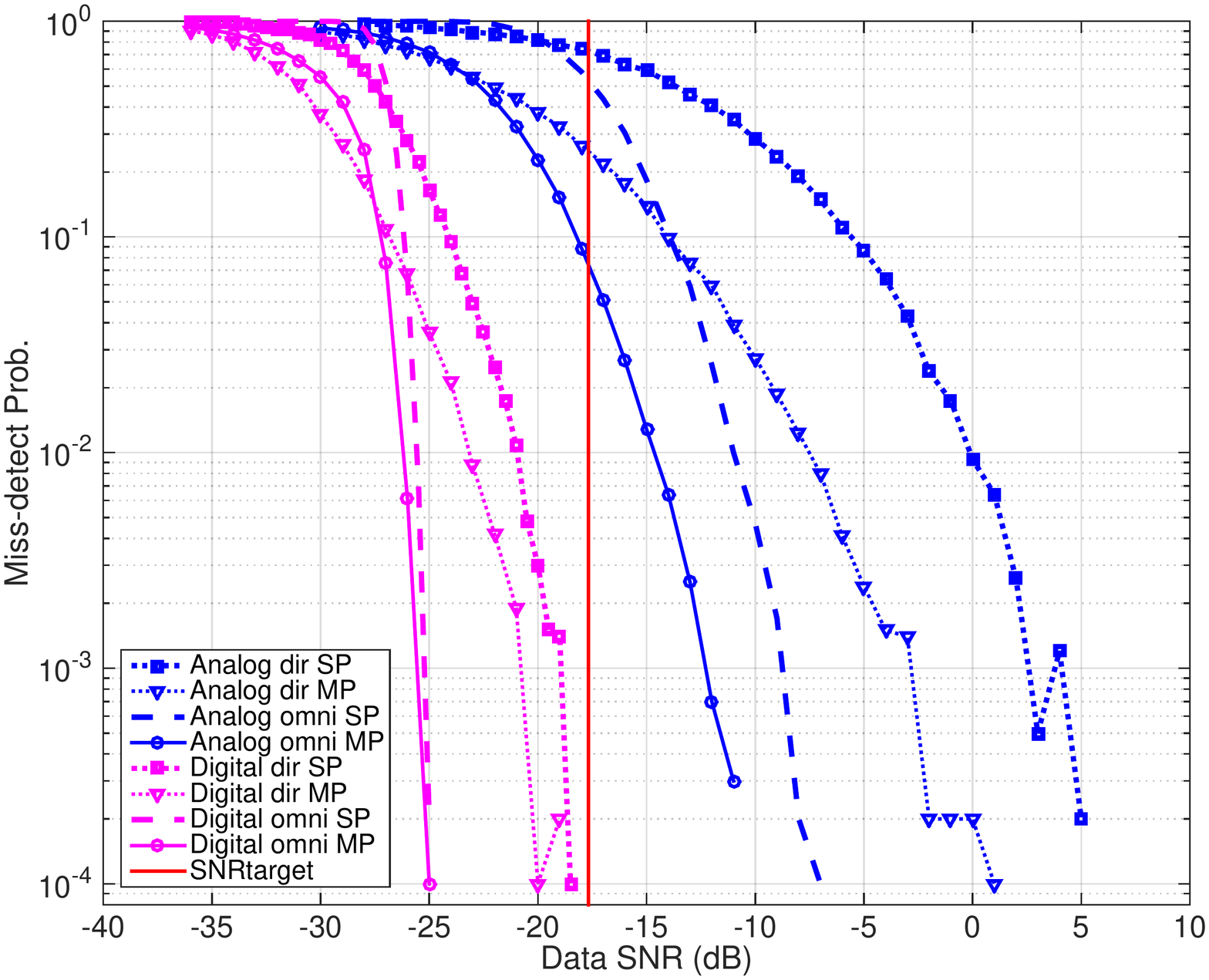}}
\caption{Misdetection Probability vs.\ data SNR: Single path channel versus multipath.}
\label{fig:NLOSLOS}
\end{figure}

The above tests were based on a theoretical single path channel.
Extensive measurements in New York City
\cite{Rappaport:12-28G,Rappaport:28NYCPenetrationLoss,Samimi:AoAD,rappaportmillimeter,McCartRapICC15}
revealed that in outdoor urban settings, the receiver can often see multiple
macro-level scattering paths with significant angular spread.
To validate the performance of our algorithm for these environments,
we tested the algorithm under a realistic statistical spatial channel model \cite{AkdenizCapacity:14}
derived from the measurements in \cite{Rappaport:12-28G,Rappaport:28NYCPenetrationLoss,Samimi:AoAD,rappaportmillimeter,McCartRapICC15}.
In this model, the channel is described by a random number of clusters, each cluster
with a random vertical and horizontal angular spread -- details are in
\cite{AkdenizCapacity:14}. Also, a more detailed statistical channel model is given in \cite{SamRapICC15}.

Fig.~\ref{fig:NLOSLOS} illustrates the comparison between the single path
vs. multipath channels.
Interestingly, both analog and digital detectors perform better in multipath than for a single path,
even though the detectors are designed based on a single-path assumption.
In the digital case the performance gain of multipath over single path is small,
while we observe a $4$~dB improvement in the analog case.
This gain is due to the fact that the multipath channel creates more opportunities
for the synchronization signal to be discovered when scanning signals at the receiver,
particularly for analog BF where the receiver can search in only one direction at a time.

\subsection{Quantization Effects}

Of course, digital beamforming comes with a high power consumption cost since the A/D power
consumption scales linearly with the number of antenna elements.  However, since power consumption also
scales exponentially with the number of bits, fully digital front ends may be feasible with very low
bit rates per antenna as proposed in \cite{Madhow:ADC,Madhow:largeArray,hassan2010analog}.
Using a linear white noise model for the quantizer \cite{GershoG:92},
it is shown in Appendix~\ref{sec:quantLoss}
that the effective SNR after quantization can be approximated as
\[
    \gamma = \frac{(1-\alpha)\gamma_0}{1-\alpha + \alpha\gamma_0},
\]
where $\gamma_0 = P/(N_0W_{tot})$ is the SNR with no quantization error,
$\gamma$ is the effective
SNR with quantization error and $\alpha$ is the average relative error of the quantizer.
Now, for a scalar uniform quantizer with only 3 bits, the relative error with optimal input level
(assuming no errors in the AGC), is $\alpha \approx 0.035$.  Using the value for $P/N_0$
at the target data rate,
the gap between $\gamma_0$ and $\gamma$ is less than 0.2 dB under our simulation assumptions.
The loss is extremely small since the SNR is already very low at the target rate, so the quantization noise
is small.
This is illustrated in Fig.~\ref{fig:DvO} where the black curves represent the performance of the digital detector with a 3-bit quantization resolution.

To give an estimation of power consumption of digital BF with $3$~bits resolution,
let us consider an A/D taken from \cite{ADCno140}, with figure of merit $P_{fm} = 59.4$~fJ per conversion step.
For a sampling frequency of twice the total bandwidth $W_{tot} = 1$~GHz, and bit resolution of $b$~bits, power consumption is defined as
\[
	P = N_r P_{fm} 2 W_{tot} 2^{b}
\]
With the number of A/Ds $N_r=16$, and $b=3$, the low-rate digital BF architecture would require only 15~mW.
We conclude that even using very low bit rates, the fully digital architecture will
not consume significant power while offering significant performance advantages
in cell search over analog BF.

\subsection{Hybrid Beamforming}
\label{hbf}

\begin{figure}
\centering {\includegraphics[width=0.8\linewidth]{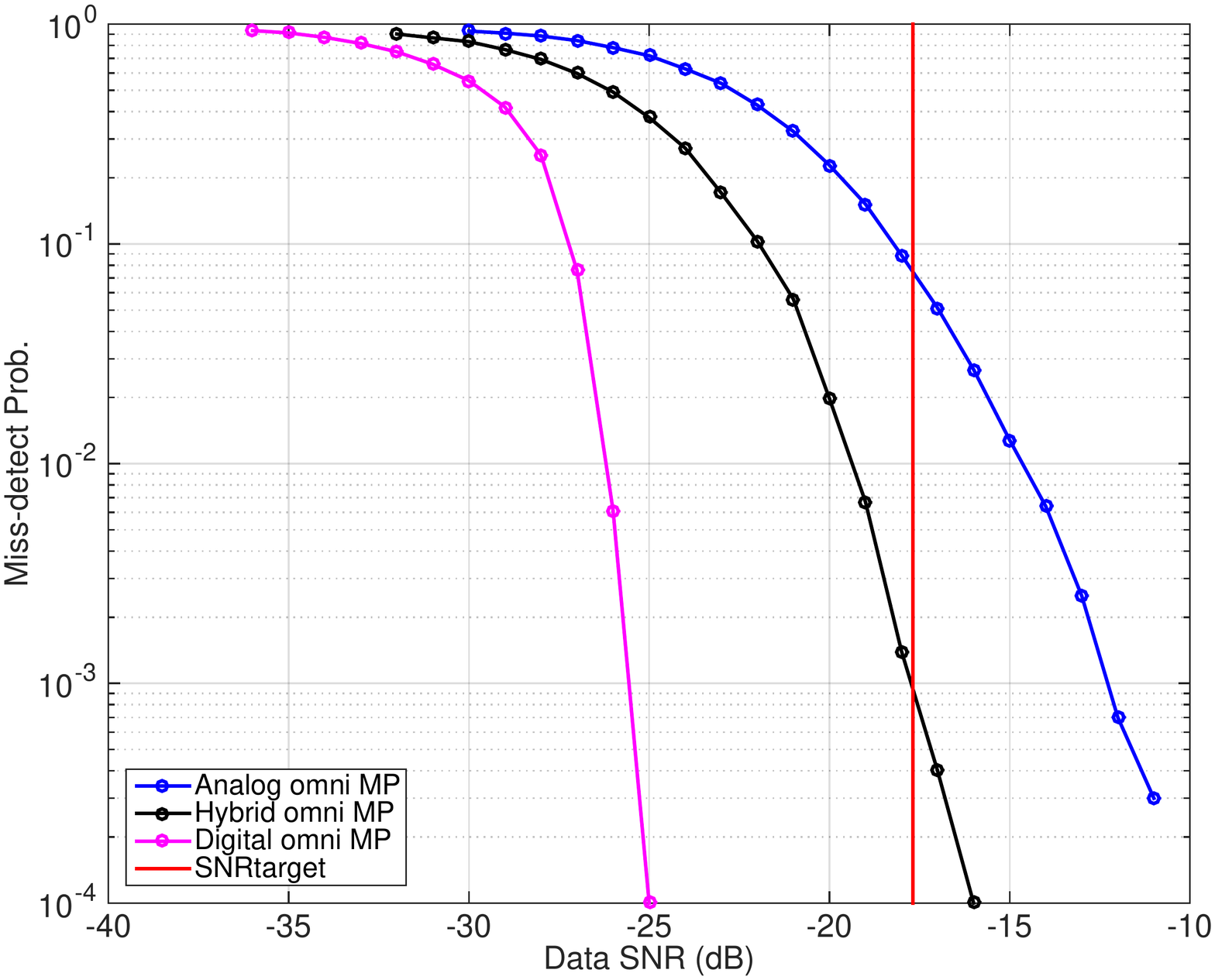}}
\caption{Misdetection Probability vs.\ data SNR: Omnidirectional transmission and multipath channel.
Hybrid beamforming's performance lies between those of analog and digital BF.}
\label{fig:hbf}
\end{figure}

Both analog and digital BF have shortcomings of their own.
The lack of multi-direction searching of the former and high power consumption of the latter have lead to a hybrid design that tries to bring a compromise between these two BF strategies
\cite{Heath:partialBF,SunRap:cm14,ghosh2014millimeter}.
With Hybrid, or quasi-analog BF, (HBF) the detector can ``look'' in more than one directions while the number of RF chains (A/Ds) are kept lower than the number of antenna elements.
Hence, power consumption is lower than digital BF.
The number of the required phase shifters, though, is as many as the size of the antenna array times the number of RF chains.
Each antenna element will receive/transmit a combination of the directional weights coming from the phase shifters connected to a different RF chain.
Due to space limitation, we refrain from going into more details regarding HBF.

We limit our comments on HBF mostly to a comparison of the performance of HBF in detecting the synchronization signal to that of analog and digital BF.
Fig. \ref{fig:hbf}, illustrates the performance of the three different detectors, in different SNR regimes.
We chose omnidirectionial transmission in multipath as our simulation scenario, since, as we showed earlier, this combination provides the best performance in both analog and digital BF.
Also, the number of RF chains in this simulation is set to $4$, resulting into $4$ receiving beams.
It can be seen that HBF's performance lies between those of analog and digital BF.
This is expected, since the main factor affecting the performance of the detector is the amount of opportunities  
it has to align the receiving beam(s) with the strongest incoming path, given a limited search time.
HBF has more chances than analog BF (it can ``look'' in more directions), but less than digital.
In this regard, HBF provides a simple method of trading off reduced search time with increased number of digital streams,
that may be useful if a fully digital architecture is not feasible.

\section*{Conclusions}
We have studied directional cell search in a mmWave cellular setting.
Two cases are presented: (i) when the base
station periodically transmits synchronization signals in random directions
to scan the angular space
and (ii) when the base station transmits signals omnidirectionally.
GLRT detectors are derived for the UE for both analog and digital beamforming.
The GLRT detectors are shown to reduce to matched filters with the synchronization signal,
with an added
eigenvector search in the digital case to locate the optimal receiver spatial signature.
Simulations were conducted under realistic parameters, both in single path channels
and multipath channels derived from actual field measurements.
The simulations indicate that omnidirectional transmission of the synchronization signal performs much better than random
angular search in both digital and analog cases.  Sequential beam searching remains to be studied.
The simulations also show that digital beamforming offers significantly
better performance than analog beamforming.
In addition, we have argued that increase in power consumption for fully digital front-ends
can be compensated by using very low bit rates per antenna with minimal loss in
performance.
These results suggest that a fully digital front-end with low bit rate per antenna
and an appropriate search algorithm may be a fundamentally
better design choice for cell search than analog beamforming.
One could thus imagine a design where low-rate digital beamforming could be used
during cell search, while analog or hybrid beamforming is used for transmissions once the directions are established.
%

\ifconf
\else

\appendices
\section{Derivation of the GLRT for Digital Beamforming} \label{sec:glrtDig}

We wish to show that, for every threshold $t'$,
the GLRT \eqref{eq:GLRTDigCts} is equivalent to a test of the form
\eqref{eq:TDigRatCts}.
To simplify the notation, throughout this section, we will fix the delay $\tau$
and drop the dependence on $\tau$ on various variables.
For example, we will write $\Lambda$ for $\Lambda(\tau)$.
In addition, it will be easier
to perform all the calculations in a finite-dimensional signal space.
To this end,
let $W_k$ be the signal space containing the signals for the $k$-th PSS time slot,
which we will assume has some finite dimension $N$ that is the same for all time slots $k$.
Since each PSS time slot
has length $T_{sig}$, we can take $N \approx W_{tot}T_{sig}$ where $W_{tot}$ is the
total signal bandwidth.  Find an orthonormal basis for each $W_k$ and for all sub-signals
in the the $k$-th time slot.  For all sub-signals $p_\ell(t)$ transmitted in the $k$-th
 let $\pbf_\ell \in \C^N$ be the sub-signal conjugate, $p_\ell^*(t)$
 in the orthonormal basis for the signal space.
Similarly, let $\Rbf_k \in \C^{N_{rx} \x N}$ be a matrix with the $N$ coefficients
of the $N_{rx}$ antenna components of the received signal vector $\ybf(t-\tau)$ in the space $W_k$.
Similarly, let $\Dbf_k \in \C^{N_{rx} \x N}$ be the matrix of the coefficients of the noise vector
$\dbf(t)$ in \eqref{eq:yrx}.  Since $\dbf(t)$ is Gaussian white noise with PSD $\nu$, the components
of $\Dbf_k$ will be white Gaussian with variance $\nu$.
With these definitions,
the two hypotheses \eqref{eq:ydigH01} can be rewritten in the finite-dimensional
signals spaces as
 \begin{subequations} \label{eq:RdigH01}
\beqa
    && H_1:  ~ \Rbf_k = \sum_{\ell \in J_k}  \alpha_\ell \ubf p_\ell^* + \Dbf_k,
        \label{eq:RdigH0} \\
    && H_0:  ~ \Rbf_k = \Dbf_k,         \label{eq:RdigH1}
\eeqa
\end{subequations}
where we recall that $J_k$ is the subset of indices $\ell$ such that the $\ell$-th subsignal
is the space $W_\ell$.  Let $\Rbf$ be the matrix of the coefficients from all $N_{slot}$
PSS time slots:
\beq \label{eq:Rdef}
    \Rbf = \left[ \Rbf_1, \cdots, \Rbf_{N_{slot}} \right].
\eeq

Since there is a one-to-one mapping between the continuous-time delay data $\Ybf_\tau$ and the
coefficients in the signal space $\Rbf$,
we can rewrite the GLRT \eqref{eq:GLRTDigCts} in terms of $\Rbf$ instead of $\Ybf_\tau$:
\beq \label{eq:GLRTDig}
    \Lambda :=  \log
        \frac{\max_{\thetabf} p(\Rbf|H_1,\thetabf)}{
        \max_{\thetabf} p(\Rbf|H_0,\thetabf) }
      \underset{H_0}{\overset{H_1}{\gtrless}} \; t'
\eeq
where
$p(\Rbf|H_i,\thetabf)$ is the density of the received data $\Rbf$
under the hypothesis $H_0$ or $H_1$ and parameters $\thetabf$.

Now, since the noise matrices
$\Dbf_k$ in \eqref{eq:RdigH01} are independent,
the log likelihood ratio \eqref{eq:GLRTDig} factors as
\beq \label{eq:LamDigSum}
    \Lambda = \min_{\nu} \sum_{k=1}^{N_{slot}} \Lambda^0_k(\nu) -
    \min_{\nu,\ubf} \sum_{k=1}^{N_{slot}} \Lambda^1_k(\nu,\ubf),
\eeq
where $\Lambda^i_k(\nu,\ubf)$ are the negative log likelihoods for the
data $\Rbf_k$ for each PSS time slot $k$:
\begin{subequations} \label{eq:LamDigk1}
\beqa
    \Lambda^0_k(\nu) &=& -\log p(\Rbf_k | H_0,\nu) \\
    \Lambda^1_k(\nu,\ubf) &=&
        \min_{\alphabf_k} \left[ -\log p(\Rbf_k | H_1,\nu,\ubf, \alphabf_k)
            \right],
\eeqa
\end{subequations}
where $\alphabf_k$ is the vector of the channel gains $\alpha_\ell$ for the sub-signals $\ell$
in the $k$-th time slot,
\[
    \alphabf_k = \left\{ \alpha_\ell, \ell \in J_k \right\}.
\]
Note that, in the $H_0$ hypothesis, there is no dependence on the parameters
$\ubf$ and $\alphabf_k$.
Now, given $\ubf$ and $\nu$, the coefficients in $\Rbf_k$ in \eqref{eq:RdigH01}
are complex Gaussians with $M := NN_{rx}$ independent components.  Hence, the likelihoods
\eqref{eq:LamDigk1} are given by
\begin{subequations} \label{eq:LamDigk2}
\beqa
    \lefteqn{ \Lambda^0_k(\nu) = \frac{1}{\nu} \|\Rbf_k\|^2_F
        + M \log(2\pi \nu) } \label{eq:LamDigH0} \\
    \lefteqn{ \Lambda^1_k(\nu,\ubf) = M \log(2\pi \nu) } \nonumber \\
    &+& \frac{1}{\nu} \min_{\alphabf_k}
        \|\Rbf_k - \sum_{\ell \in J_k} \alpha_\ell \ubf\pbf_\ell^* \|^2_F  \label{eq:LamDigH1a}.
\eeqa
\end{subequations}
Since we have assumed that the different sub-signals $p_\ell(t)$ are orthogonal,
the vector representations $\pbf_\ell$ will also be orthogonal.  Hence, the minimization
in \eqref{eq:LamDigH1a} is given by
\beqa
 \lefteqn{ \Lambda^1_k(\nu,\ubf) = M \log(2\pi \nu) } \nonumber \\
    &+& \frac{1}{\nu} \left[ \|\Rbf_\ell \|^2_F - \sum_{\ell \in J_k}
        \frac{|\ubf^*\Rbf_k \pbf_\ell|^2}
        {\|\pbf_\ell\|^2\|\ubf\|^2} \right].
            \label{eq:LamDigH1b}
\eeqa

We next compute the minimizations over $\nu$ and $\ubf$.
For the $H_0$ hypothesis, we apply \eqref{eq:LamDigH0} to obtain
\beqa
    \lefteqn{ \min_{\nu} \sum_{k=1}^{N_{slot}} \Lambda^0_k(\nu)
    = \min_{\nu} \sum_{k=1}^{N_{slot}} \left[
    \frac{1}{\nu} \|\Rbf_k\|^2_F
        + M \log(2\pi \nu) \right] } \nonumber \\
    &=& \min_{\nu} \left[
    \frac{1}{\nu} \|\Rbf\|^2_F + N_{slot}M \log(2\pi \nu) \right] \nonumber \\
    &=& N_{slot}M + N_{slot}M \log\left[ \frac{2\pi}{N_{slot}M} \|\Rbf\|^2_F \right], \label{eq:LamDigH0min}
    \hspace{2cm}
\eeqa
where we have used the fact that
\beq \label{eq:RnormSum}
    \|\Rbf\|^2_F = \sum_{k=1}^{N_{slot}} \|\Rbf_k\|^2_F.
\eeq

For the $H_1$ hypothesis, first observe that
\beqa
    \lefteqn{ \sum_{k=1}^{N_{slot}}\left\{
        \|\Rbf_k\|^2_F - \sum_{\ell \in J_k} \frac{|\ubf^*\Rbf_k \pbf_\ell|^2}
        {\|\pbf_\ell\|^2\|\ubf\|^2} \right\} } \nonumber \\
        &\stackrel{(a)}{=}&  \sum_{k=1}^{N_{slot}}  \|\Rbf_k\|^2_F  - \sum_{\ell =1}^L
        \frac{|\ubf^*\Rbf_{\sigma(\ell)} \pbf_\ell|^2}
        {\|\pbf_\ell\|^2\|\ubf\|^2} \nonumber \\
        &\stackrel{(b)}{=}& \|\Rbf\|^2_F - \frac{\|\ubf^*\Vbf\|^2}{\|\ubf\|^2},
        \label{eq:LamDigH1min1}
\eeqa
where in step (a), we have used the notation that $\sigma(\ell)$ is the index $k$ for the PSS time-slot
in which the sub-signal $p_\ell(t)$ is transmitted.  In step (b), we have again used \eqref{eq:RnormSum}
and re-written the sum over $\ell$ using a matrix $\Vbf$ defined as
\beq \label{eq:Vdef1}
    \Vbf = \left[ \frac{1}{\|\pbf_1\|}\Rbf_{\sigma(1)}\pbf_1, \cdots,
    \frac{1}{\|\pbf_L\|}\Rbf_{\sigma(L)}\pbf_L \right].
\eeq
The minimum of \eqref{eq:LamDigH1min1}
over $\ubf$ is given by the maximum singular value \cite{GolubV:89}:
\beq \label{eq:LamDigH1min2}
    \min_{\ubf} \|\Rbf\|^2_F - \frac{\|\ubf^*\Vbf\|^2}{\|\ubf\|^2}
    = \|\Rbf\|^2_F - \sigma_{max}^2(\Vbf),
\eeq
Substituting \eqref{eq:LamDigH1min1} and \eqref{eq:LamDigH1min2}
into \eqref{eq:LamDigH1b} we obtain
\ifonecol
\beqa
    \lefteqn{ \min_{\nu,\ubf} \sum_{\ell=1}^L \Lambda^1_\ell(\nu,\ubf) }
    \nonumber \\
    &=& \min_{\nu} \frac{1}{\nu} \left[
     \|\Rbf\|^2_F - \sigma_{max}^2(\Vbf) \right] + N_{slot}M \log(2\pi \nu) \nonumber \\
    &=& N_{slot}M \nonumber
     + N_{slot}M\log\left[ \frac{2\pi}{N_{slot}M}(  \|\Rbf\|^2_F - \sigma_{max}^2(\Vbf) )
        \right]. \label{eq:LamDigH1min3}
\eeqa
\else
\beqa
    \lefteqn{ \min_{\nu,\ubf} \sum_{\ell=1}^L \Lambda^1_\ell(\nu,\ubf) }
    \nonumber \\
    &=& \min_{\nu} \frac{1}{\nu} \left[
     \|\Rbf\|^2_F - \sigma_{max}^2(\Vbf) \right] + N_{slot}M \log(2\pi \nu) \nonumber \\
    &=& N_{slot}M \nonumber  \\
    && + N_{slot}M\log\left[ \frac{2\pi}{N_{slot}M}(  \|\Rbf\|^2_F - \sigma_{max}^2(\Vbf) )
        \right]. \label{eq:LamDigH1min3}
\eeqa
\fi
Substituting \eqref{eq:LamDigH0min} and \eqref{eq:LamDigH1min3} into
\eqref{eq:LamDigSum}, we see that the GLRT is given by
\beqa
    \lefteqn{ \Lambda = -N_{slot}M \log \left[ 
         1 - \frac{\sigma^2_{max}(\Vbf)}{\|\Rbf\|^2_F} \right] } \nonumber \\
    &=& -N_{slot}M \log \left[ (1 - T) \right],\label{eq:GLRDig1}
\eeqa
where $T$ is the statistic,
\beq \label{eq:TDigRat}
    T := \frac{\sigma^2_{max}(\Vbf)}{\|\Rbf\|^2_F} .
\eeq
Since $\Lambda$ is an increasing function of the statistic $T$,
a ratio test \eqref{eq:GLRTDig} is equivalent to a test of the form
\[
    T \underset{H_0}{\overset{H_1}{\gtrless}} \; t.
\]

It remains to show that the statistic $T$ in \eqref{eq:TDigRat} is identical to $T(\tau)$ in
\eqref{eq:TDigRatCts}.  To this end, first recall that the matrices $\Rbf_k$ are the
coefficients of the received signal $\ybf(t-\tau)$ in the signal space for the
time slot $I_k$.  Since the coefficients
are in an orthonormal basis, the energy is preserved so that
\[
    \|\Rbf_k\|^2_F = \int_{t \in I_k+\tau} \|\ybf(t)\|^2dt =
    \int_{I_k} \|\ybf(t-\tau)\|^2dt.
\]
Therefore, the energy $E(\tau)$ in \eqref{eq:EdigCts} is given by
\beq \label{eq:EeqDig}
   E(\tau) = \sum_{k=1}^{N_{slot}} \|\Rbf_k\|^2_F = \|\Rbf\|_F^2.
\eeq
Similarly, for every $\ell \in J_k$,
$\pbf_\ell$ is the vector representation of the sub-signal $p_\ell^*(t)$ in the signal space
for the $k$-th PSS time slot. Since we are using an orthonormal basis,
$\Rbf_k\pbf_\ell$ must be inner product,
\[
    \Rbf_k\pbf_\ell = \int_{I_k} \ybf(t-\tau)p_\ell^*(t)dt = \int \ybf(t-\tau)p_\ell^*(t)dt,
\]
where the last step is valid since the support of $p_\ell(t)$ is contained in the interval $I_k$.
Therefore, the normalized correlation $\vbf_\ell(\tau)$ is precisely
\[
    \vbf(\tau) = \frac{1}{\|\pbf_\ell\|}\Rbf_k\pbf_\ell.
\]
This identity implies that the matrix $\Vbf(\tau)$ in \eqref{eq:VdefCts} is precisely \eqref{eq:Vdef1}.
Combining this fact with \eqref{eq:EeqDig} shows that
statistic $T$ in \eqref{eq:TDigRat} is identical to $T(\tau)$ in
\eqref{eq:TDigRatCts}.  We conclude that the GLRT \eqref{eq:GLRTDigCts} is equivalent to
the correlation test \eqref{eq:TDigRatCts}.

\section{Derivation of the GLRT for Analog Beamforming} \label{sec:glrtAnalog}

Similar to the previous section, we prove that for every threshold $t'$, the GLRT
\eqref{eq:GLRTAnalogCts} is equivalent to a test of the form \eqref{eq:TRatAnalogCts}.
Again, the
dependence on $\tau$ is dropped to simplify the notation.
As before, we use a finite dimensional signal space representation for all the signals.
Let $\rbf_k \in \C^N$ be the vector representation of the delayed
beamformed signal $z(t-\tau)$ in the signal space $W_k$ for the $k$-th PSS time slot.
Similarly, let $\dbf_k$ and $\pbf_\ell \in \C^N$ be the representation for the noise signal $d(t)$
and PSS sub-signal $p_\ell(t)$.
Then, the two hypotheses \eqref{eq:zH01} can be re-written as
\begin{subequations} \label{eq:ranaH01}
\beqa
    && H_1:  ~ \rbf_k = \sum_{\ell \in J_k}  \beta_\ell \pbf_\ell + \dbf_k,
        \label{eq:ranaH0} \\
    && H_0:  ~ \rbf_k = \dbf_k,         \label{eq:ranaH1}
\eeqa
\end{subequations}
Let $\Rbf$ be the matrix of the data for all the time slots
\[
    \Rbf = \left[ \rbf_1, \cdots, \rbf_{N_{slot}} \right].
\]
The GLRT \eqref{eq:GLRTAnalogCts} can then be re-written as
\beq \label{eq:GLRTAna}
    \Lambda :=  \log
        \frac{\max_{\thetabf} p(\Rbf|H_1,\thetabf)}{
        \max_{\thetabf} p(\Rbf|H_0,\thetabf) }
      \underset{H_0}{\overset{H_1}{\gtrless}} \; t'
\eeq
where
$p(\Rbf|H_i,\thetabf)$ is the density of the received data $\Rbf$
under the hypothesis $H_0$ or $H_1$ and parameters $\thetabf$.

As before, the  likelihood ratio in \eqref{eq:GLRTAna} factors as:
\beq \label{eq:LamAnSum}
    \Lambda = \min_{\nu} \sum_{k=1}^{N_{slot}} \Lambda^0_k(\nu) -
    \min_{\nu} \sum_{k=1}^{N_{slot}} \Lambda^1_k(\nu),
\eeq
where $\Lambda^i_k(\nu)$ are the negative log likelihoods for the
data $\rbf_k$ for each sub-signal $k$:
\begin{subequations} \label{eq:LamAnak1}
\beqa
    \Lambda^0_k(\nu) &=& -\log p(\rbf_k | H_0,\nu) \\
    \Lambda^1_k(\nu) &=&
        \min_{\betabf_k} \left[ -\log p(\rbf_\ell | H_1,\nu,\betabf_k)
            \right],
\eeqa
\end{subequations}
where $\betabf_k$ is the vector of channel gains for the $k$-th time slot
\[
    \betabf_k = \left\{ \beta_\ell~|~ \ell \in J_k \right\}.
\]
We know that the noise vector $\dbf_k$ in \eqref{eq:ranaH01}
is a vector of length $N$ consisting of independent complex gaussian random variables.
Hence, using the hypothesis models \eqref{eq:ranaH01},
the likelihoods in \eqref{eq:LamAnak1} are equal to:
\begin{subequations} \label{eq:LamAnak2}
\beqa
    \lefteqn{ \Lambda^0_k(\nu) = \frac{1}{\nu} \|\rbf_k\|^2
        + N \log(2\pi \nu) } \label{eq:LamAnH0} \\
    \lefteqn{ \Lambda^1_k(\nu) = N\log(2\pi \nu) } \nonumber \\
    &+& \frac{1}{\nu} \min_{\betabf_k}
        \|\rbf_k - \sum_{\ell \in J_k} \beta_\ell\pbf_\ell \|^2. \label{eq:LamAnH1min1}
\eeqa
\end{subequations}
Since the vectors $\pbf_\ell$ are orthogonal, the minimization in \eqref{eq:LamAnH1min1} is given by
\beqa
    \lefteqn{ \Lambda^1_k(\nu) = N\log(2\pi \nu) } \nonumber
\\
    &+& \frac{1}{\nu} \left[
     \|\rbf_k\|^2  - \sum_{\ell \in J_k} \frac{|\rbf_k^*\pbf_\ell|^2}{\|\pbf_\ell\|^2} \right].
        \label{eq:LamAnH1min2}
\eeqa

Next, we perform the minimizations of both hypotheses over $\nu$. For the $H_0$ hypothesis, we have:
\beqa
    \lefteqn{ \min_{\nu} \sum_{k=1}^{N_{slot}} \Lambda^0_k(\nu)
    = \min_{\nu} \sum_{k=1}^{N_{slot}} \left[
     \frac{1}{\nu} \|\rbf_k\|^2
        + N\log(2\pi \nu) \right] } \nonumber \\
    &=& \min_{\nu} \left[
    \frac{1}{\nu} \|\rbf\|^2 + NN_{slot} \log(2\pi \nu) \right] \nonumber \\
    &=& NN_{slot} + NN_{slot} \log\left[ \frac{2\pi}{NN_{slot}} \|\rbf\|^2 \right], \label{eq:LamAnH0min}
    \hspace{2cm}
\eeqa
where we have used the fact that
\beq \label{eq:rnormSumAna}
    \|\rbf\|^2 = \sum_{k=1}^{N_{slot}} \|\rbf_k\|^2.
\eeq
For the $H_1$ case, we first evaluate the summation over the slots,
\beqa
    \lefteqn{ \sum_{k=1}^{N_{slot}}\left\{ \|\rbf_k\|^2 -
        \sum_{\ell \in J_k} \frac{|\rbf_k^*\pbf_\ell|^2}{\|\pbf_\ell\|^2} \right\} } \nonumber \\
        &\stackrel{(a)}{=}&  \sum_{k=1}^{N_{slot}}  \|\rbf_k\|^2  - \sum_{\ell =1}^L
        \frac{|\rbf_{\sigma(\ell)}^* \pbf_\ell|^2}
        {\|\pbf_\ell\|^2 } \nonumber \\
        &\stackrel{(b)}{=}& \|\rbf\|^2 - \|\vbf\|^2,
        \label{eq:LamAnaH1min3}
\eeqa
where in step (a), as before,
we have used the notation that $\sigma(\ell)$ is the index $k$ for the PSS time-slot
in which the sub-signal $p_\ell(t)$ is transmitted.  In step (b), we have again used \eqref{eq:rnormSumAna}
and re-written the sum over $\ell$ using the vector $\vbf$ defined as
\beq \label{eq:vdefana}
    \vbf = \left[ \frac{1}{\|\pbf_1\|}\rbf_{\sigma(1)}^*\pbf_1, \cdots,
    \frac{1}{\|\pbf_L\|}\rbf_{\sigma(L)}^*\pbf_L \right].
\eeq
Combining \eqref{eq:LamAnH1min2} and \eqref{eq:LamAnaH1min3}, we obtain:
\beqa
    \lefteqn{ \min_{\nu}  \sum_{k=1}^{N_{slot}} \Lambda^1_k(\nu)  }   \nonumber \\
    &=& \min_{\nu} \left\{ NN_{slot} \log(2\pi \nu) +
        \frac{1}{\nu}(\|\rbf\|^2 - \|\vbf\|^2) \right\} \nonumber \\
    &=& NN_{slot} + NN_{slot}\log\left[ \frac{2\pi}{NN_{slot}}(  \|\rbf\|^2 - \|\vbf\|^2 )
        \right]. \label{eq:LamAnH1min4}
\eeqa
Substituting \eqref{eq:LamAnH1min4} and \eqref{eq:LamAnH0min} into \eqref{eq:LamAnSum}, the GLRT is given by:
\beqa \label{eq:GLRAn1}
    \lefteqn{\Lambda = -NN_{slot} \log \left[
        \left( 1 - \frac{\|\vbf\|^2}{\|\rbf\|^2} \right) \right]}
	\nonumber \\
        &=& -NN_{slot} \log \left[ (1 - T) \right],
\eeqa
where $T$ is the statistic,
\beq \label{eq:Tana}
    T = \frac{\|\vbf\|^2}{\|\rbf\|^2}.
\eeq
As before, we can conclude that since $\Lambda$ is an increasing function of $T$,
for every threshold level $t'$,
the ratio test given by \eqref{eq:GLRTAna} is equivalent to a test of the form
\beq \label{eq:TRatAnalog}
    T \underset{H_0}{\overset{H_1}{\gtrless}} \; t,
\eeq
for some threshold level $t$.  With a similar argument to the digital beamforming case,
we can show that the $T$ defined in \eqref{eq:Tana} is precisely the test statistic
$T(\tau)$ in \eqref{eq:TRatAnalogCts}.  We conclude that the GLRT in \eqref{eq:GLRTAnalogCts}
is equivalent to a correlation threshold test \eqref{eq:TRatAnalogCts}.

\section{Simulation Parameter Selection Details}\label{sec:simParam}
In this section,
we provide more details on the logic behind the selection of their simulation parameters, and
also illustrate some of the considerations that should be made in selecting these values
for practical systems.

\paragraph{Signal parameters}
As mentioned in Section~\ref{sec:design}, the synchronization
signal is divided into $N_{sig}$ narrow-band sub-signals sent over different
frequency bands to provide frequency diversity.
Our experiments indicated that $N_{sig}=4$ provides a good
tradeoff between frequency diversity and coherent combining.
However, due to space, we do not report the results for other values
of $N_{sig}$.
For the length of the PSS interval, we took $T_{sig}=$ 100 $\mu s$,
which is sufficiently small that the channel will be coherent even at the
very high frequencies for mmWave.  For example, at a UE with velocity of $30$ kmph
and 28 GHz,
the maximum Doppler shift of the mmWave channel is $\approx 780$ Hz, so the coherence time
is $\gg 100~\mu$s.
Measurements of typical delay spreads in a mmWave outdoor setting indicate delay spreads within
a narrow angular region to be typically less $< 30\; \text{ns}$ \cite{Rappaport:12-28G, McCartRapICC15}.
This means that if we take the sub-signal bandwidth of $W_{sig}$ = 1 MHz, the channel
will be relatively flat across this band.

Now, since the PSS signals occur every $T_{per}$ seconds,
the PSS overhead will be $T_{sig}/T_{per}$. With our parameter selection, the overhead of the signal is $2$\%.
In fact, any $T_{per}$ greater than $5$~ms gives us an overhead less than 2\%.

\paragraph{False alarm target} To set an appropriate false alarm target,
we recognize that false alarms on the PSS result in additional
searches for a secondary synchronization signal (SSS),
which costs both computational power and increases the false alarm rate for the SSS.
The precise acceptable
level for the PSS false alarm rate will depend on the SSS signal design.
In this simulation, we will assume that the total false alarm rate is at most
$R_{FA}$ = 0.01 false alarm per search period.  Thus, the false alarm rate per delay hypothesis
must be $P_{FA} \leq R_{FA}/(N_{hyp})$ where $N_{hyp}$ is the number
of hypotheses that will be tested per second.  The number of hypotheses
are given by the product
\[
    N_{hyp} = N_{dly}N_{PSS}N_{FO},
\]
where $N_{dly}$ is the number of delay hypotheses per transmission period $T_{per}$,
$N_{PSS}$ is the number of PSS waveforms that can be transmitted by the
base station and $N_{FO}$ is the number of frequency offsets.
The values for these will depend on whether the UE is performing the cell
search for initial access or while in connected mode for handover.
In this paper, we consider only the initial access case.

For initial access, the UE must search over delays in the range $\tau \in [0,T_{per}]$.
Assuming the correlations are computed
at twice the bandwidth, we will have
$N_{dly} = 2W_{sig} T_{per} = 2 (10)^6 \times 5 (10)^{-3} = 10^4$.
For the number of PSS signals, we will take $N_{PSS} = 3$, which is same as the number of signals offered by the current LTE system.
However, we may need to increase this number to accommodate more cell IDs in a dense BS deployment.
To estimate the number of frequency offsets, suppose that the initial
frequency offset can be as much as 1 ppm at 28~GHz.  This will result in
an frequency offset of $28(10)^3$ Hz.
The maximum Doppler shift will add approximately 780~Hz, giving a total
initial frequency offset error of $\Delta f_{max}=28 + 0.78$~kHz.
In order that the channel does not rotate
more than 90$^\circ$ over the period of $T_{sig}=100$ $\mu$s, we need
a frequency accuracy of $\Delta f= 10^{4}/4$.  Since $2\Delta f_{max}/\Delta f = 23.0$,
it will suffice to take $N_{FO}=23$ frequency offset hypotheses.  Hence, the target
FA rate for initial access should be
\[
    P_{FA} = \frac{R_{FA}}{N_{hyp}} = \frac{0.01}{10^4 (3) 23} = 1.4493(10)^{-8}.
\]

\paragraph{Antenna pattern} We assume a set of two dimensional antenna arrays at both the BS and the UE.
On the BS side, the array is comprised of $8 \times 8$ elements and on the receiver side we have $4 \times 4$ elements.
The spacing of the elements is set at $\lambda /2$, where $\lambda$ is the wavelength.
These antenna patterns were considered in \cite{AkdenizCapacity:14} and showed to offer excellent
system capacity for small cell urban deployments.
In addition, a $4 \times 4$ array operating in the
$28 \; \text{GHz}$ band,
for instance, will have dimensions of roughly $1.5\; \text{cm} \times 1.5 \; \text{cm}$.\\

\section{Quantization Effects} \label{sec:quantLoss}
To estimate the effect of quantization, we use a standard AWGN model for the quantization noise
\cite{GershoG:92}.
Let $r[n]$ be the complex samples of some signal that we model as a random process of the form,
\beq \label{eq:rxw}
    r[n] = x[n] + w[n], \quad E|x[n]|^2 = E_s, ~ E|w[n]|^2 = N_0,
\eeq
where $x[n]$ represents the signal, $w[n]$ is the noise, and
$E_s$ and $N_0$ are the signal and noise energy per sample.
The SNR of this signal is
\[
    \gamma_0 = E_s/N_0.
\]
Now consider a quantized version of this signal, $\rhat[n] = Q(r[n])$
where $Q(\cdot)$ is a scalar quantizer applied to each sample.  It is shown in
\cite{FletcherRGR:07} that the quantizer can be modeled as
\beq \label{eq:rhatQ}
    \rhat[n] = Q(r[n]) = (1-\alpha)r[n] + v[n],
\eeq
where $v[n]$ is quanitization noise that is uncorrelated with $r[n]$
and has variance
\beq \label{eq:vQ}
    E|v[n]|^2 = \alpha(1-\alpha) E|r[n]|^2 = \alpha(1-\alpha)(E_s + N_0).
\eeq
In \eqref{eq:rhatQ} and \eqref{eq:vQ}, $\alpha$ is the relative quantization error
(or inverse coding gain using the terminology in \cite{GershoG:92}),
\[
    \alpha := \frac{E|Q(r)-r|^2}{E|r|^2}.
\]
This relative error is a function of the number of bits of the quantizer, the
choice of the quantizer levels and the distribution of the quantizer input.
Table~\ref{tbl:quant} shows the values for $\alpha$ (in dB scale) for
various bit values for a simple scalar uniform quantizer with Gaussian input.
For the values in the table, a simple numerical optimization is run to optimize the step size
to obtain the minimum relative error $\alpha$.  In practice, the actual signal level will not
arrive at the optimal level due to imperfections in the A/D -- so a few more bits are likely needed.

Substituting \eqref{eq:rxw} into \eqref{eq:rhatQ}, we obtain
\[
     \rhat[n] = Q(r[n]) = (1-\alpha)x[n] + (1-\alpha)w[n] + v[n].
\]
Hence, the effective SNR after quantization is
\ifonecol
\beqan
    \lefteqn{ \gamma = \frac{(1-\alpha)^2E|x[n]|^2}{
        (1-\alpha)^2E|w[n]|^2 + E|v[n]|^2} } \nonumber\\
    &=& \frac{(1-\alpha)^2E_s}{(1-\alpha)^2N_0 + \alpha(1-\alpha)(E_s + N_0) }
    = \frac{(1-\alpha)\gamma_0}{1 + \alpha \gamma_0 }.
\eeqan
\else
\beqan
    \lefteqn{ \gamma = \frac{(1-\alpha)^2E|x[n]|^2}{
        (1-\alpha)^2E|w[n]|^2 + E|v[n]|^2} } \nonumber\\
    &=& \frac{(1-\alpha)^2E_s}{(1-\alpha)^2N_0 + \alpha(1-\alpha)(E_s + N_0) } \nonumber \\
    &=& \frac{(1-\alpha)\gamma_0}{1 + \alpha \gamma_0 }.
\eeqan
\fi
For the PSS detection, suppose that the PSS signal arrives at power $P$ and the noise has
a power spectral density $N_0$.  Since the PSS arrives in $N_{sig}$ sub-signals
each having bandwidth $W_{sig}$, the energy per orthogonal sample will be
\[
    E_s = P / (W_{sig}N_{sig})
\]
and hence the SNR will be
\[
    \gamma_0 = P/N_0 = P/(W_{sig}N_{sig}N_0).
\]

\begin{table}
\centering
\begin{tabular}{|p{2cm}|p{4cm}|}
  \hline
  {\bf Number bits} & {\bf Coding gain, $-10\log_{10}(\alpha)$ (dB)}
  \tabularnewline \hline
  1 &   4.4 \tabularnewline \hline
  2 &   9.3 \tabularnewline \hline
  3 &   14.5 \tabularnewline \hline
\end{tabular}
\caption{Coding gain values as a function of the number of bits for
a scalar uniform quantizer with Gaussian noise and an optimized step size.
\label{tbl:quant} }
\end{table}
\fi

\bibliographystyle{IEEEtran}
\bibliography{bibl}

\end{document}